\documentclass[aps,prb,times,twocolumn,superscriptaddress,showpacs,a4paper]{revtex4-1}
\usepackage{colortbl,marvosym}
\usepackage{amsfonts,amsmath,amssymb}
\usepackage{graphicx,wasysym}
\usepackage{tabularx,hhline}
\usepackage[dvipsnames]{xcolor}
\usepackage{color}

\makeatletter
\newcommand*{\rom}[1]{\expandafter\@slowromancap\romannumeral #1@}
\makeatother

\begin{document}
%%%%%%%%%%%%%%%%%%%%%%%%%%%%%%%%%%%%%%%%%%%%%%

\title{Reentrance of disorder in the anisotropic shuriken Ising model}

\author{Rico Pohle}
\affiliation{Okinawa Institute of Science and Technology Graduate University,
Onna-son, Okinawa 904-0495, Japan}

\author{Owen Benton}
\affiliation{Okinawa Institute of Science and Technology Graduate University,
Onna-son, Okinawa 904-0495, Japan}

\author{L.D.C. Jaubert}
\affiliation{Okinawa Institute of Science and Technology Graduate University,
Onna-son, Okinawa 904-0495, Japan}

\date{\today}
\begin{abstract}
For a material to order upon cooling is common sense. What is more seldom is for disorder to reappear at lower temperature, which is known as reentrant behavior. Such resurgence of disorder has been observed in a variety of systems, ranging from Rochelle salts to nematic phases in liquid crystals. Frustration is often a key ingredient for reentrance mechanisms. Here we shall study a frustrated model, namely the anisotropic shuriken lattice, which offers a natural setting to explore an extension of the notion of reentrance between magnetic \textit{disordered} phases. By tuning the anisotropy of the lattice, we open a window in the phase diagram where magnetic disorder prevails down to zero temperature. In this region, the competition between multiple disordered ground states gives rise to a double crossover where both the low- and high-temperature regimes are less correlated than the intervening classical spin liquid. This reentrance of disorder is characterized by an entropy plateau, a multi-step Curie law crossover and a rather complex diffuse scattering in the static structure factor. Those results are confirmed by complementary numerical and analytical methods: Monte Carlo simulations, Husimi-tree calculations and an exact decoration-iteration transformation.
\end{abstract}
\maketitle

%%%%%%%%%%%%%%%%%%%%%%%%%%%%%%%%%%%%%%%%%
%%%%%%%%%%%%%%%%%%%%%%%%%%%%%%%%%%%%%%%%%
%%%%%%%%%%%%%%%%%%%%%%%%%%%%%%%%%%%%%%%%%

Recent progress in frustrated magnetism has delivered entire maps of long-range ordered and disordered phases, obtained for example via the variation of bond anisotropy~\cite{Savary12a,Lee12a,Hao14a,Chernyshev14a,Gotze15a,Oitmaa16a,Essafi16a} or further nearest-neighbour couplings~\cite{Messio12a,He15a,He15b,Bieri15a,Iqbal15a,McClarty15a,Henelius16a}. Such phase diagrams have allowed to put a series of frustrated materials onto a global and connected map, that can be experimentally explored via physical or chemical pressure~\cite{Zhou12a,Dun14a,Wiebe15a,Rau15a}. On such phase diagrams, when two ordered phases meet, an enhancement of the classical ground-state degeneracy takes place~\cite{Yan13a}. This degeneracy can either be lifted by thermal fluctuations, giving rise to multiple phase transitions~\cite{Jaubert15b,Robert15a}, or may destroy any kind of order down to (theoretically) zero temperature. This is where spin liquids appear. But this picture is less clear at the frontier between ordered and (possibly multiple) disordered ground states. In particular how do disordered phases compete with each other at finite temperature ?

\begin{figure}[t]
\centering\includegraphics[width=0.45\textwidth]{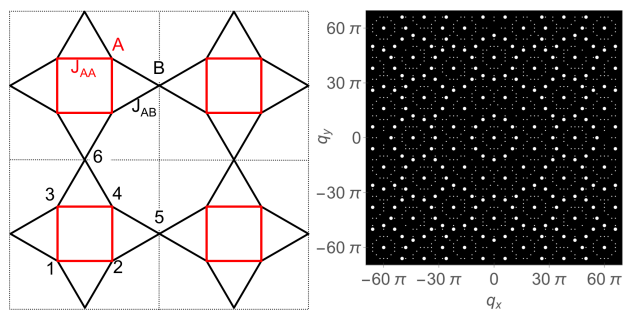}
\caption{{\bf The shuriken lattice} as seen in real space (\textit{left}) and Fourier space (\textit{right}). There are 6 sites per unit-cell with two sublattices $A$ and $B$. Interactions between A-sites (square plaquettes) are described with coupling constant $J_{AA}$ (red), while interactions between A- and B-sites (octagonal plaquettes) are described with $J_{AB}$ (black).}
\label{fig:UnitCell}	
\end{figure}

The frustrated shuriken lattice~\cite{Nakano2013} -- also known as square-kagome~\cite{Siddharthan2001,Richter2004,Derzhko2006,Richter2009,Derzhko2013,Nakano2014,Nakano2015,Ralko2015}, squagome~\cite{Tomczak03a,Glaetzle14a}, squa-kagome~\cite{Rousochatzakis2013} or L4-L8~\cite{Rousochatzakis2013} lattice -- provides an interesting model-example for such competition. Being made of corner-sharing triangles, it is locally similar to the famous kagome lattice, but with the important difference that the shuriken lattice is composed of two inequivalent sublattices [see Fig.~\ref{fig:UnitCell}]. Such asymmetry offers a natural setup for lattice anisotropy. In the asymptotic limits of this anisotropy, a promising zero-temperature phase diagram has emerged for quantum spin$-1/2$, ranging from a bipartite long-range ordered phase to a highly degenerate ground state made of tetramer clusters of spins~\cite{Rousochatzakis2013}. However, while the quantum ground states~\cite{Richter2009,Nakano2013,Rousochatzakis2013,Ralko2015} and the influence of a magnetic field~\cite{Richter2004,Derzhko2006,Richter2009,Derzhko2013,Derzhko14a,Nakano2013,Nakano2014,Nakano2015} have been studied to some extent, little is known about the finite-temperature properties in zero field~\cite{Siddharthan2001,Tomczak03a}.\\

%%%%%%%%%%%%%%%%%%%%%
%%%%%%%%%%%%%%%%%%%%%

%%%%%%%%%%%%%%%%%%%%%
\begin{figure*}[t]
\centering
\includegraphics[width=\textwidth]{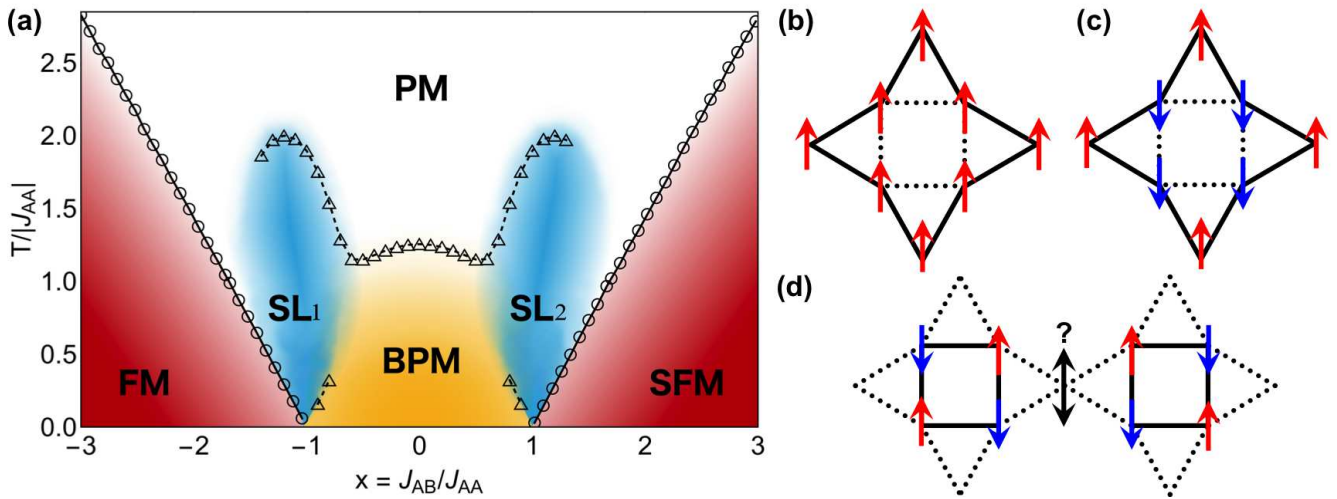}
\caption{
{\bf Phase Diagram of the Ising model on the anisotropic shuriken lattice}. (a) The circles (triangles) correspond to phase transitions (crossovers), obtained by Monte Carlo simulations (Husimi-tree calculations) [see appendix~\ref{sec:Methods} for further details]. As a function of the coupling ratio $x = \frac{J_{AB}}{J_{AA}}$, the model supports a long-range ordered ferromagnet (FM) [see panel (b)], a long-range ordered staggered ferromagnet (SFM) [see panel (c)], a binary paramagnet (BPM) [see panel (d)], and two classical spin liquids (SL$_{1,2}$). The BPM illustrated in panel (d) is made of antiferromagnetically ordered square plaquettes, decoupled from each other and from the intermediate spins sitting on the B sublattice. For $|x|\gtrsim 1$, on cooling, the system undergoes a evolution from ``gas $\xrightarrow[]{\rm crossover}$ liquid $\xrightarrow[]{\rm transition}$ solid''. As for $|x|\lesssim 1$, it provides a remarkable example of reentrance from ``gas $\xrightarrow[]{\rm crossover}$ liquid $\xrightarrow[]{\rm crossover}$ gas''.
}
\label{fig:PhaseDiagram}	
\end{figure*}
%%%%%%%%%%%%%%%%%%%%%

In this paper, our goal is to develop a comprehensive and precise understanding of the frustrated phase diagram of the Ising model on the anisotropic shuriken lattice, relying on a combination of numerical and analytical methods (Monte Carlo simulations, Husimi tree calculations and decoration-iteration transformation). Using the lattice anisotropy as a tuning parameter, we find that this model supports two long-range ordered phases (ferromagnet (FM) and staggered ferromagnet (SFM)), two classical spin liquids (SL$_{1,2}$) characterized by complex static structure factors, and a \textit{zero-temperature} paramagnet composed of two kinds of isolated (super)spins with strictly zero correlations between them. We shall refer to this latter phase as a \textit{binary paramagnet}. Over an extended region of the phase diagram, there is a double crossover from the high-temperature paramagnet to the spin liquids and finally into the low-temperature binary paramagnet. This double crossover gives rise to a non-monotonic behavior of the correlation length, which can be seen as an analogue of reentrant behavior between disordered phases. As a by-product, we notice an essentially perfect agreement between Husimi-tree analytics and Monte Carlo simulations in the disordered regimes.\\

The paper is divided as follows. The model is introduced in section~\ref{sec:Model}, followed by its phase diagram in section~\ref{sec:PhaseDiagram}. In section~\ref{sec:LiquidRegime}, we analyze in details the double-crossover region between disordered regimes. We conclude the paper by discussing possible experimental realizations of the shuriken lattice and summarizing our results in sections~\ref{sec:exp} and \ref{sec:conclusion} respectively. Most technical details are given in the appendices.

%%%%%%%%%%%%%%%%%%%%%%%%%%%%%%%%%%%%%%%%%
%%%%%%%%%%%%%%%%%%%%%%%%%%%%%%%%%%%%%%%%%
%%%%%%%%%%%%%%%%%%%%%%%%%%%%%%%%%%%%%%%%%
\section{Anisotropic shuriken model}
\label{sec:Model}

The shuriken lattice is made of corner-sharing triangles with 6 sites per unit cell [see Fig.~\ref{fig:UnitCell}]. As opposed to its kagome parent where all spins belong to hexagonal loops, the shuriken lattice forms two kinds of loops made of either 4 or 8 sites. As a consequence, $2/3$ of the spins in the system belong to the A-sublattice, while the remaining $1/3$ of the spins form the B-sublattice. Let us respectively define $J_{AA}$ and $J_{AB}$ as the coupling constants between A-sites on the square plaquettes, and between A- and B-sites on the octagonal plaquettes. The Hamiltonian of the model can be written as:
\begin{equation}
H = - J_{AA}\;\sum_{\langle i j \rangle_{AA}} \sigma^{A}_i \sigma^{A}_j
- J_{AB}\;\sum_{\langle i j \rangle_{AB}} \sigma^{A}_i \sigma^{B}_j
\label{eq:ham}
\end{equation}
where we consider Ising spins $\sigma_{i}=\pm 1$ with nearest-neighbor coupling.

There is no frustration for ferromagnetic $J_{AA} = +1$ where the system undergoes a phase transition with spontaneous $\mathbb{Z}_{2}$ symmetry breaking for $J_{AB}\neq 0$. We shall thus focus on antiferromagnetic $J_{AA} = -1$, which will be our energy and temperature scale of reference. The thermodynamics will be discussed as a function of the coupling ratio~[\onlinecite{Derzhko2006},\onlinecite{Rousochatzakis2013},\onlinecite{Ralko2015}]
\begin{equation}
x = \dfrac{J_{AB}}{J_{AA}},
\label{eq:coupling}
\end{equation}
with ferro- and antiferromagnetic $J_{AB}$.

%%%%%%%%%%%%%%%%%%%%%%%%%%%%%%%%%%%%%%%%%
%%%%%%%%%%%%%%%%%%%%%%%%%%%%%%%%%%%%%%%%%
%%%%%%%%%%%%%%%%%%%%%%%%%%%%%%%%%%%%%%%%%
\section{Phase Diagram}
\label{sec:PhaseDiagram}

The Hamiltonian of equation~(\ref{eq:ham}) is invariant under the transformation
\begin{align}
\sigma^{A}\rightarrow -\sigma^{A}
\nonumber\\
J_{AB} \rightarrow - J_{AB}
\label{eq:symmetricPD}
\end{align}
All quantities derived from the energy, and especially the specific heat $C_{h}$ and entropy $S$, are thus the same for $x$ and $-x$. Their respective magnetic phases are related by reversing all spins of the A-sublattices.

%%%%%%%%%%%%%%%%%%%%%%%%%%%%%%%%%%%%%%%%%
\subsection{Long-range order: ${|x|>1}$}
\label{sec:LRO}

When the octagonal plaquettes are dominating ($x\rightarrow \pm \infty $), the shuriken lattice becomes a decorated square lattice, with A-sites sitting on the bonds between B-sites. Being bipartite, the decorated square lattice is not frustrated and orders via a phase transition of the 2D Ising Universality class~\cite{Sun06a} by spontaneous $\mathbb{Z}_{2}$ symmetry breaking. Non-universal quantities such as the transition temperature can be exactly computed by using the decoration-iteration transformation~\cite{Fisher59a,Sun06a,Strecka15b} [see appendix~\ref{subsec:exactTc}]
\begin{eqnarray}
T_{c}=\frac{2 J_{AB}}{\ln\left(\sqrt{2}+1 +\sqrt{2+2\sqrt{2}}\right)}
\approx 1.30841 J_{AB}
\label{eq:exactTc}
\end{eqnarray}
The low-temperature ordered phases, displayed in Fig.~\ref{fig:PhaseDiagram}.(b) and~\ref{fig:PhaseDiagram}.(c), remain the ground states of the anisotropic shuriken model for $x<-1$ and $x>1$ respectively. The persistence of the 2D Ising Universality class down to $|x|=1^{+}$ is not necessarily obvious, but is confirmed by finite-size scaling from Monte Carlo simulations [see appendix~\ref{sec:PhaseTransition}].

These two ordered phases are respectively ferromagnetic (FM, $x<-1$) and staggered ferromagnetic (SFM, $x>1$) [see Fig.~\ref{fig:PhaseDiagram}.(b,c)]. The staggering of the latter comes from all spins on square plaquettes pointing in one direction, while the remaining ones point the other way. This leads to the rather uncommon consequence that fully antiferromagnetic couplings -- both $J_{AA}$ and $J_{AB}$ are negative for $x>1$ -- induce long-range ordered (staggered) ferromagnetism, reminiscent of Lieb ferrimagnetism~\cite{Lieb89a} as pointed out in Ref.~[\onlinecite{Rousochatzakis2013}] for quantum spins. The existence of ferromagnetic states among the set of ground states of Ising antiferromagnets is not rare, with the triangular and kagome lattices being two famous examples. But such ferromagnetic states are usually part of a degenerate ensemble where no magnetic order prevails on average. Here the lattice anisotropy is able to induce ferromagnetic order in an antiferromagnetic model by lifting its ground-state degeneracy at $|x|=1$ (see below). This is interestingly quite the opposite of what happens in the spin-ice model~\cite{Harris97a}, where frustration prevents magnetic order in a ferromagnetic model by stabilizing a highly degenerate ground state.

%%%%%%%%%%%%%%%%%%%%%
\begin{figure*}[t]
\centering\includegraphics[width=1 \textwidth]{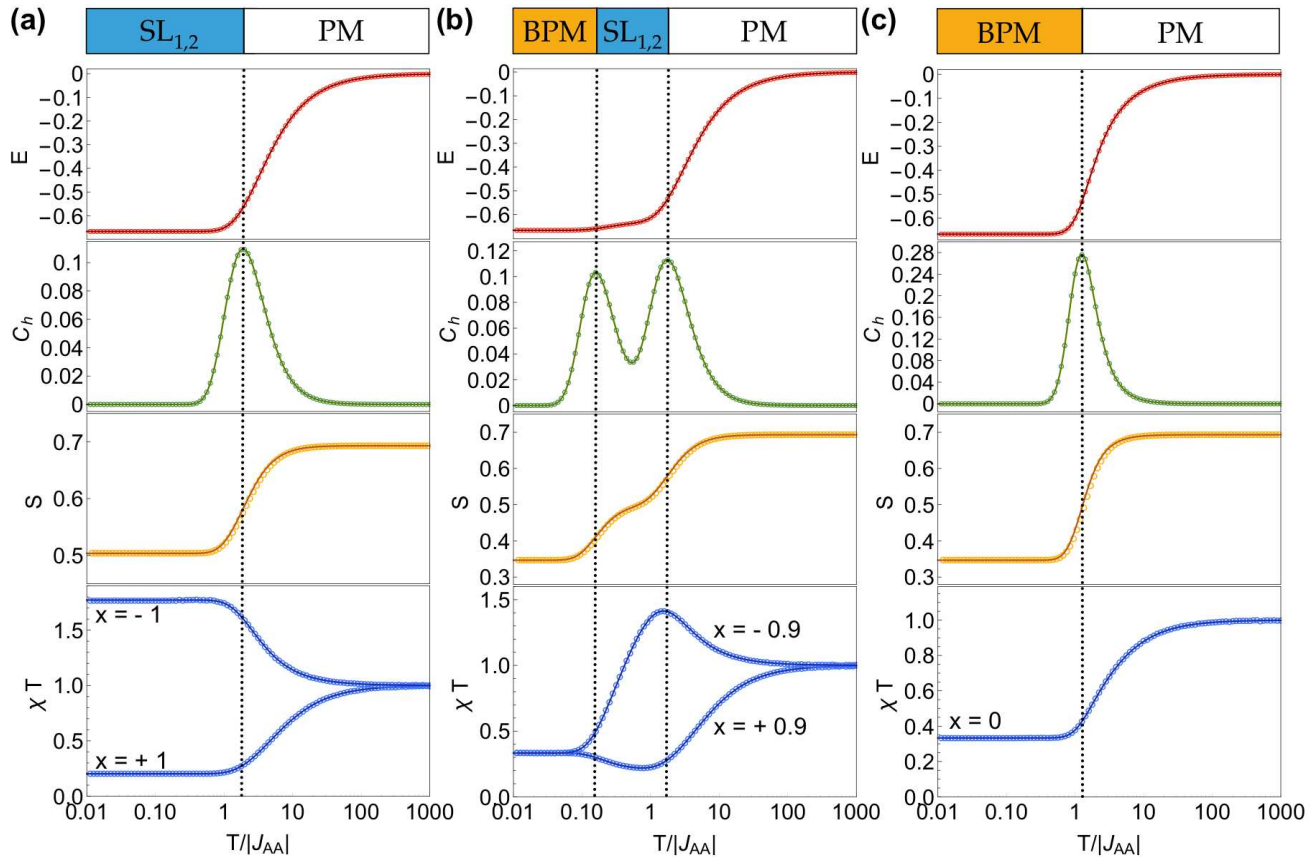}
\caption{{\bf Multiple crossovers between the paramagnetic, spin-liquids and binary regimes} as observed in the specific heat $C_h$, entropy $S$ and reduced magnetic susceptibility $\chi \, T$. The models correspond to a) $x = \pm1$, b) $x = \pm 0.9$ and c) $x = 0$. There is no phase transition for this set of parameters, which is why the Husimi tree calculations (lines) perfectly match the Monte Carlo simulations (circles) for all temperatures. The double crossover is present for $x = \pm 0.9$, with the low-temperature regime being the same as for $x=0$, as confirmed by its entropy and susceptibility. The entropy is obtained by integration of $C_{h}/T$, setting $S(T\rightarrow +\infty) = \ln 2$. The vertical dashed lines represent estimates of the crossover temperatures determined by the local specific-heat maxima. The temperature axis is on a logarithmic scale. All quantities are given per number of spins and the Boltzmann constant $k_{B}$ is set to 1.
}
\label{fig:MCHT}
\end{figure*}
%%%%%%%%%%%%%%%%%%%%%
%%%%%%%%%%%%%%%%%%%%%%%%%%%%%%%%%%%%%%%%%
\subsection{Binary paramagnet: ${|x|<1}$}

The central part of the phase diagram is dominated by the square plaquettes. The ground states are the same for all $|x|<1$. A sample configuration of these ground states is given in Fig.~\ref{fig:PhaseDiagram}.(d), where antiferromagnetically ordered square-plaquettes are separated from each other via spins on sublattice B. The antiferromagnetic square-plaquettes locally order in two different configurations equivalent to a superspin $\Xi$ with Ising degree of freedom.
\begin{eqnarray}
\Xi=\sigma^{A}_{1}-\sigma^{A}_{2}-\sigma^{A}_{3}+\sigma^{A}_{4}=\pm 4,
\label{eq:superspin}
\end{eqnarray}
where the site indices are given in Fig.~\ref{fig:UnitCell}. These superspins are the classical analogue of the tetramer objects observed in the spin$-1/2$ model~\cite{Rousochatzakis2013}. At zero temperature, the frustration of the $J_{AB}$ bonds perfectly decouples the superspins $\Xi$ from the B-sites. The system can then be seen as two interpenetrating square lattices: one made of superspins, the other one of B-sites. We shall refer to this phase as a \textit{binary} paramagnet (BPM).

The perfect absence of correlations beyond square plaquettes at $T=0$ allows for a simple determination of the thermodynamics. Let $N_{uc}$ and $N=6\,N_{uc}$ be respectively the total number of unit cells and spins in the system, and $\langle X \rangle$ be the statistical average of $X$. There are $N_{uc}$ square plaquettes and $2N_{uc}$ B-sites, giving rise to an extensive ground-state entropy
\begin{eqnarray}
S_{\rm BPM}=k_{B}\,\ln\left(2^{N_{uc}}\,2^{2N_{uc}}\right)=\dfrac{N}{2}k_{B}\,\ln 2
\label{eq:entropyBPM}
\end{eqnarray}
which turns out to be half the entropy of an Ising paramagnet. As for the magnetic susceptibility $\chi$, it diverges as $T\to 0^{+}$. But the reduced susceptibility $\chi\, T$, which is nothing less than the normalized variance of the magnetization
\begin{eqnarray}
\chi\,T &=& \dfrac{1}{N}\left(\sum_{i,j}\langle \sigma_{i}\sigma_{j}\rangle\,-\,\langle \sigma_{i}\rangle\langle \sigma_{j}\rangle\right),\nonumber\\
&=&1+\dfrac{1}{N}\sum_{i\neq j}\langle \sigma_{i}\sigma_{j}\rangle,
\label{eq:Chi}
\end{eqnarray}
converges to a finite value in the BPM
\begin{eqnarray}
\chi\,T|_{\rm BPM}&=&\dfrac{1}{3}.
\label{eq:bulkChi}
\end{eqnarray}
%

%%%%%%%%%%%%%%%%%%%%%%%%%%%%%%%%%%%%%%%%%
\subsection{Classical spin liquid: ${|x|\sim 1}$}

There is a sharp increase of the ground-state degeneracy at $|x|=1$, when the binary paramagnet and the (staggered) ferromagnet meet. As is common for isotropic triangle-based Ising antiferromagnets, 6 out of 8 possible configurations per triangle minimize the energy of the system. As opposed to the BPM one does not expect a cutoff of the correlations [see section~\ref{sec:corr}], making these phases cooperative paramagnets~\cite{Villain79a}, also known as classical spin liquids.

Due to the high entropy of these cooperative paramagnets, the SL$_{1,2}$ phases spread to the neighboring region of the phase diagram for $|x|\sim 1$ and $T>0$, continuously connected to the high-temperature paramagnet [see Fig.~\ref{fig:PhaseDiagram}]. Hence, for $|x|\gtrsim 1$, the anisotropic shuriken model stabilizes a cooperative paramagnet above a non-degenerate\footnote{besides the trivial time-reversal symmetry} long-range ordered phase. This is a general property of classical spin liquids when adiabatically tuned away from their high-degeneracy point, as observed for example in Heisenberg antiferromagnets on the kagome~\cite{Elhajal02a} or pyrochlore~\cite{Canals08a,Chern10b,Mcclarty14b} lattices, and possibly in the material of Er$_{2}$Sn$_{2}$O$_{7}$~\cite{Yan13a}. For $|x|\lesssim 1$ on the other hand, multiple crossovers take place upon cooling which deserves a dedicated discussion in the following section~\ref{sec:LiquidRegime}.

%%%%%%%%%%%%%%%%%%%%%%%%%%%%%%%%%%%%%%%%%
%%%%%%%%%%%%%%%%%%%%%%%%%%%%%%%%%%%%%%%%%
%%%%%%%%%%%%%%%%%%%%%%%%%%%%%%%%%%%%%%%%%
\section{Reentrance of disorder}
\label{sec:LiquidRegime}

%%%%%%%%%%%%%%%%%%%%%%%%%%%%%%%%%%%%%%%%%
\subsection{Double crossover}
\label{sec:dblecross}

First of all, panels (a) and (c) of Fig.~\ref{fig:MCHT} confirm that the classical spin liquids and binary paramagnet persist down to zero temperature for $x=\pm 1$ and $x=0$ respectively, and that all models for $|x|\leqslant 1$ have extensively degenerate ground states. For $x=\pm 0.9$ there is a double crossover indicated by the double peaks in the specific heat $C_h$ of Fig.~\ref{fig:MCHT}.(b). These peaks are not due to phase transitions since they do not diverge with system size. The double crossover persists for $0.5\lesssim |x|<1$. Upon cooling, the system first evolves from the standard paramagnet to a spin liquid before entering the binary paramagnet. The intervening spin liquid takes the form of an entropy plateau for $|x|=0.9$ [see Fig.~\ref{fig:MCHT}.(b)], at the same value as the low-temperature regime for $|x|=1$ [see Fig.~\ref{fig:MCHT}.(a)]. All relevant thermodynamic quantities are summarized in Table~\ref{tab:zeroT}.

\begin{figure}[t]
\centering\includegraphics[width=0.5 \textwidth]{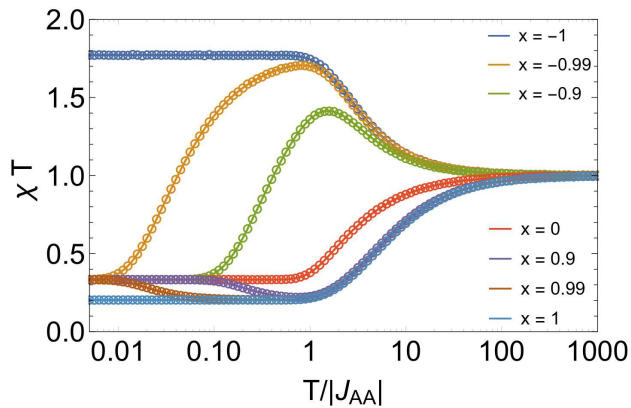}
\caption{{\bf Reduced susceptibility $\chi\, T$} with coupling ratios of $x = \pm 1,  \pm 0.99, \pm 0.9$ and $0$, obtained from Husimi-tree calculations (solid lines) and Monte Carlo simulations (circles). The Curie-law crossover of classical spin liquids is standard, \textit{i.e.} $\chi\, T$ is monotonic, for $x=\pm 1$ and 0, and takes a multi-step behavior for intermediate values of $x$, due to the double crossover. The characteristic values of the entropy and reduced susceptibility are given in Table~\ref{tab:zeroT}. The temperature axis is on a logarithmic scale
}
\label{fig:Chi_All}	
\end{figure}

While the mapping of equation~(\ref{eq:symmetricPD}) ensures the invariance of the energy, specific heat and entropy upon reversing $x$ to $-x$, it does not protect the magnetic susceptibility. The build up of correlations in classical spin liquids is known to give rise to a Curie-law crossover~\cite{Jaubert13a} between two $1/T$ asymptotic regimes of the susceptibility, as observed in pyrochlore~\cite{Isakov04a,Ryzhkin05a,Conlon10a,Jaubert13a}, triangular~\cite{Isoda08a} and kagome~\cite{Isoda08a,Li10a,Macdonald11a} systems. This is also what is observed here on the anisotropic shuriken lattice for $x=\{-1,0,1\}$ [see Fig.~\ref{fig:Chi_All}]. But for intermediate models with $x=\{-0.99,-0.9,0.9,0.99\}$, the double crossover makes the reduced susceptibility non-monotonic. $\chi\, T$ first evolves towards the values of the spin liquids SL$_{1}$ (resp. SL$_{2}$) for $x<0$ (resp. $x>0$) before converging to $1/3$ in the binary paramagnet.

Beyond the present problem on the shuriken lattice, this multi-step Curie-law crossover underlines the usefulness of the reduced susceptibility to spot intermediate regimes, and thus the proximity of different phases. From the point of view of renormalization group theory, the $(x,T)=(\pm 1,0)$ coordinates of the phase diagram are fixed points which deform the renormalization flows passing in the vicinity.

%
%%%%%%%%%%%%%%%%%%%%%
\newcolumntype{C}{>{}c<{}}
\renewcommand{\arraystretch}{3}
\begin{table}
\centering
\begin{tabular}{||C|C|C|C||}
\hhline{|t:====:t|}
$T\to0^{+}$& Monte Carlo & Husimi tree & $\quad$ exact $\quad$ \\
\hhline{||====||}
$S (|x| = 1)$ & $ 0.504(1) $ & $ \dfrac{1}{6} \ln \dfrac{41}{2} \approx 0.5034$ & n/a \\
\hhline{||----||}
$\chi \, T (x = 1)$ & $ 0.203(1) $ & $0.2028$ & n/a \\
\hhline{||----||}
$\chi \, T (x =-1)$ & $ 1.766(1) $ & $1.771$ & n/a \\
\hhline{||====||}
$S (|x| < 1)$ & $ 0.347(1) $ & $ \dfrac{1}{2} \ln 2 \approx 0.3466$ & $ \dfrac{1}{2} \ln 2 $ \\
\hhline{||----||}
$\chi \, T (|x| < 1)$ & $ 0.333(1) $ & $\dfrac{1}{3}$ &$ \  \dfrac{1}{3} $ \\
\hhline{|b:====:b|}
\end{tabular}
\caption{{\bf Entropies $S$ and reduced susceptibilities $\chi T$ as $\mathbf{T\to0^{+}}$} for the anisotropic shuriken lattice with coupling ratios $|x|\leqslant 1$. The results are obtained from Monte Carlo simulations, Husimi tree analytics and the exact solution for the binary paramagnet. All quantities are given per number of spins and the Boltzmann constant $k_{B}$ is set to 1.
}		
\label{tab:zeroT}
\end{table}
%%%%%%%%%%%%%%%%%%%%%
%%%%%%%%%%%%%%%%%%%%%
%%%%%%%%%%%%%%%%%%%%%

%%%%%%%%%%%%%%%%%%%%%%%%%%%%%%%%%%%%%%%%%
\subsection{Decoration-iteration transformation}
\label{section:checkerboard}

The phase diagram of the anisotropic shuriken model and, in particular, the double crossover observed for $|x| < 1$ [see Fig.~\ref{fig:PhaseDiagram}] can be further understood using an exact mapping to an effective model on the checkerboard lattice, a method known as decoration-iteration transformation [see Ref.~[\onlinecite{Strecka15b}] for a review]. In short, by summing over the degrees of freedom of the A-spins, one can arrive at an effective Hamiltonian involving only the B-spins, which form a checkerboard lattice. The coupling constants of the effective Hamiltonian are functions of the temperature $T$ and for  $|x| < 1$ they vanish at both high and low temperatures, but are finite for an intermediate regime. This intermediate regime may be identified as the SL$_{1,2}$ cooperative paramagnets of Fig.~\ref{fig:PhaseDiagram}, whereas the low-temperature region of vanishing effective interaction corresponds to the binary paramagnet (BPM). This mapping is able to predict a non-monotonic behavior of the correlation length.

In this section we give a brief sketch of the derivation of the effective model, before turning to its results. Details of the effective model are given in Appendix \ref{appendix:mapping}.

To begin, consider the partition function for the system, with the Hamiltonian given by Eq.~(\ref{eq:ham})
\begin{eqnarray}
Z=
\sum_{\{ \sigma^{A}_i=\pm1\}}
\sum_{\{ \sigma^{B}_i=\pm1\}}
\exp\left[
-\beta H
\right]
\label{eq:partition0}
\end{eqnarray}
where $\beta=\frac{1}{T}$ is the inverse temperature and the sums are over all possible spin configurations. Since in the Hamiltonian of Eq.~(\ref{eq:ham}) the square plaquettes of the A-sites are only connected to each other via their interaction with the intervening B-sites, it is possible to directly take the sum over configurations of A-spins in Eq.~(\ref{eq:partition0}) for a fixed (but completely general) configuration of B-spins. Doing so, we arrive at
\begin{eqnarray}
Z=
\sum_{\{ \sigma^{B}_i=\pm1\}}
\prod_{\square}
\mathcal{Z}_{\square} (\{ \sigma^{B}_i\})
\label{eq:partitionSq}
\end{eqnarray}
where the product is over all the square plaquettes of the lattice and $\mathcal{Z}_{\square} (\{ \sigma^{B}_i\})$ is a function of the four B-spins immediately neighbouring a given square plaquette. The B-spins form a checkerboard lattice, and  Eq.~(\ref{eq:partitionSq}) can be exactly rewritten in terms of an effective Hamiltonian $H_{\boxtimes}$ on that lattice:
\begin{eqnarray}
&&Z=
\sum_{\{ \sigma^{B}_i=\pm1\}}
\exp(-\beta \sum_{\boxtimes} H_{\boxtimes}) \\
&&H_{\boxtimes}= 
-\mathcal{J}_0(T)-
\mathcal{J}_1(T) \sum_{\langle ij \rangle}\sigma^{B}_i \sigma^{B}_j + \nonumber \\
&& \qquad
-
\mathcal{J}_2(T) \sum_{\langle \langle ij \rangle \rangle}\sigma^{B}_i \sigma^{B}_j
-
\mathcal{J}_{\sf ring}(T) \prod_{i \in \boxtimes} \sigma_i^B
\label{eq:effectiveHcheck}
\end{eqnarray}
where $\sum_{\boxtimes}$ is a sum over checkerboard plaquettes of B-spins. The effective Hamiltonian $H_{\boxtimes}$ contains a constant term $\mathcal{J}_0$, a nearest neighbour interaction $\mathcal{J}_1$, a second nearest neighbour interaction $\mathcal{J}_2$, and a four-site ring interaction $\mathcal{J}_{\sf ring}$. All couplings are functions of temperature $\mathcal{J}_i=\mathcal{J}_i(T)$ and are invariant under the transformation $J_{AB}\longmapsto -J_{AB}$ because the degrees of freedom of the A-sites have been integrated out. Expressions for the dependence of the couplings on temperature are given in Appendix \ref{appendix:mapping}. 

The temperature dependence of the effective couplings  $\mathcal{J}_i=\mathcal{J}_i(T)$ can itself give rather a lot of information about the behavior of the shuriken model.

%%%%%%%%%%%%%%%%%%%%%
\begin{figure}
\centering
\includegraphics[width=0.9\columnwidth]{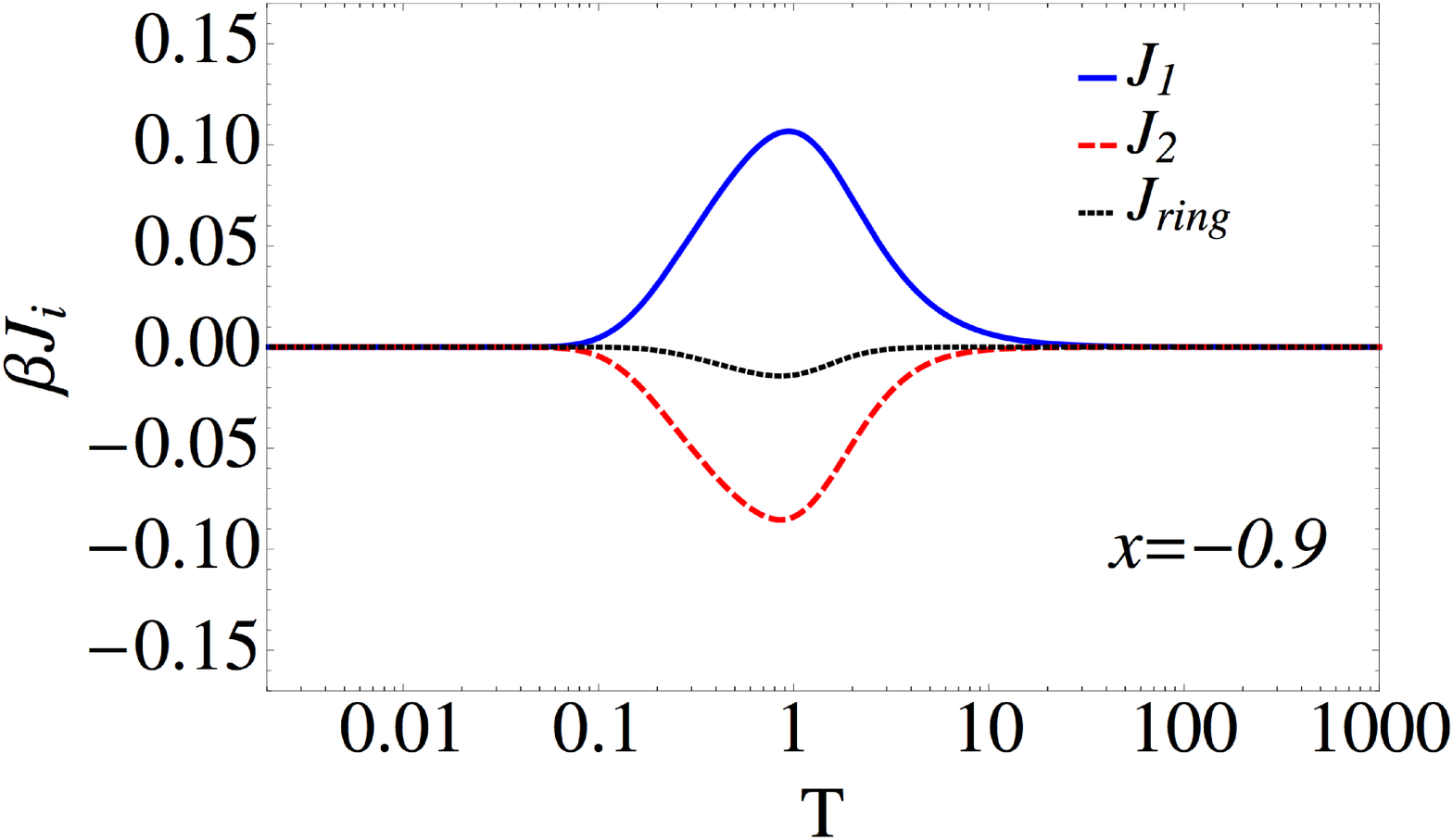}\\ 
\includegraphics[width=0.9\columnwidth]{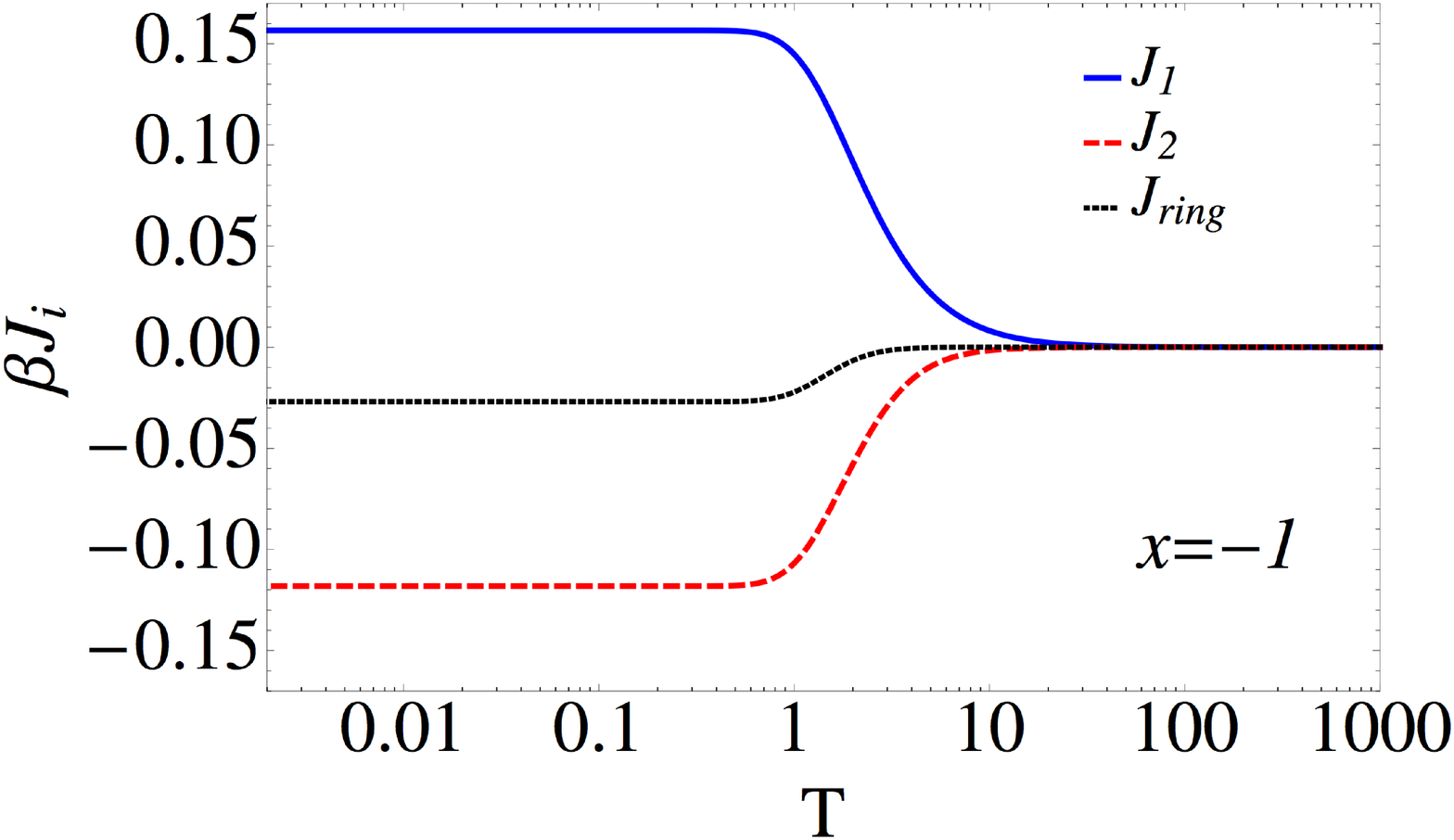}
\caption{
\textbf{Behavior of the coupling constants of the the effective checkerboard lattice model as a function of temperature} [Eq.~(\ref{eq:effectiveHcheck})] for $x=-0.9$ (upper panel) and $x=-1$ (lower panel). {\it Upper panel}: All couplings vanish at both high and low temperatures with an intermediate regime at $T \sim 1$ where the effective interactions are stronger. The intermediate regime corresponds to the spin liquid region of the phase diagram Fig.~\ref{fig:PhaseDiagram}, with the high- and low-temperature regimes corresponding to the paramagnet and binary paramagnet respectively. {\it Lower panel}: For all couplings $\mathcal{J}_i$, $\beta \mathcal{J}_i$ vanishes at high temperature and tends to a finite constant of magnitude $|\beta \mathcal{J}_i (T)|<<1$ at low temperature. The short range correlated, spin-liquid regime, thus extends all the way down to $T=0$. 
}
\label{fig:couplings}
\end{figure}
%%%%%%%%%%%%%%%%%%%%%

%%%%%%%%%%%%%%%%%%%%%
\begin{figure}
\centering
\includegraphics[width=0.9\columnwidth]{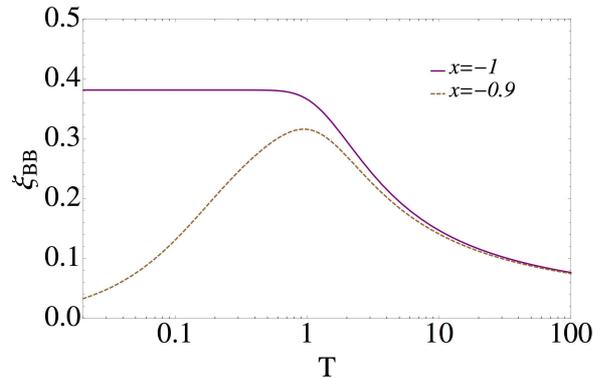}% 
\caption{
\textbf{Correlation lengths in the effective checkerboard model}, calculated from Eq.~(\ref{eq:corrlen}), for $x=-0.9$ and $x=-1$. The correlation length is calculated to leading order in a perturbative expansion of the effective model in powers of $\beta \mathcal{J}_i$. Such an expansion is reasonable for $|x|\leq1$ since $\beta \mathcal{J}_i<<1$ for all $T$ (see Fig.~\ref{fig:couplings}). For $x=- 0.9$ the behavior of the correlation length is non-monotonic. The correlation length is maximal in the spin liquid regime but correlations remain short ranged at all temperatures. In the binary paramagnet regime, the correlation length vanishes linearly at low temperature. For $x=- 1$, the correlation length enters a plateau at $T\sim 1$, and short range correlations remain down to $T=0$.
}
\label{fig:correlationlengths}
\end{figure}
%%%%%%%%%%%%%%%%%%%%%

First we consider the case $|x|<1$. In this regime of parameter space, all effective interactions $\mathcal{J}_1, \mathcal{J}_2, \mathcal{J}_{\sf ring}$ vanish exponentially at low temperature $T<<1$. For intermediate temperatures $T \sim 1$ the effective interactions in Eq.~(\ref{eq:effectiveHcheck}) become appreciable before vanishing once more at high temperatures. This is illustrated for the case $x=-0.9$ in the upper panel of Fig.~\ref{fig:couplings}. Seeing the problem in terms of these effective couplings gives some intuition into the double crossover observed in simulations. As the temperature is decreased the effective couplings $|\mathcal{J}_i|$ increase in absolute value and the system enters a short range correlated regime. However, as the temperature decreases further, the antiferromagnetic correlations on the square plaquettes of A-spins become close to perfect, and act to screen the effective interaction between B-spins. This is reflected in the exponential suppression of the couplings $\mathcal{J}_1, \mathcal{J}_2, \mathcal{J}_{\sf ring}$.

In the case $|x|=1$, the effective interactions $\mathcal{J}_i$ no longer vanish exponentially at low temperature, but instead vanish linearly
\begin{eqnarray}
\mathcal{J}_1, \mathcal{J}_2, \mathcal{J}_{\sf ring} \sim T.
\end{eqnarray}
The ratio of effective couplings to the temperature $\beta \mathcal{J}_{i}$ thus tends to a constant below $T \sim 1$, as shown in the lower panel of Fig.~\ref{fig:couplings}. Thus, the zero temperature limit of the shuriken model can be mapped to a finite temperature model on the checkerboard lattice for $|x|=1$ and to an infinite temperature model for $|x|<1$.\\

%%%%%%%%%%%%%%%%%%%%%
\begin{figure*}
\centering\includegraphics[width=0.9\textwidth]{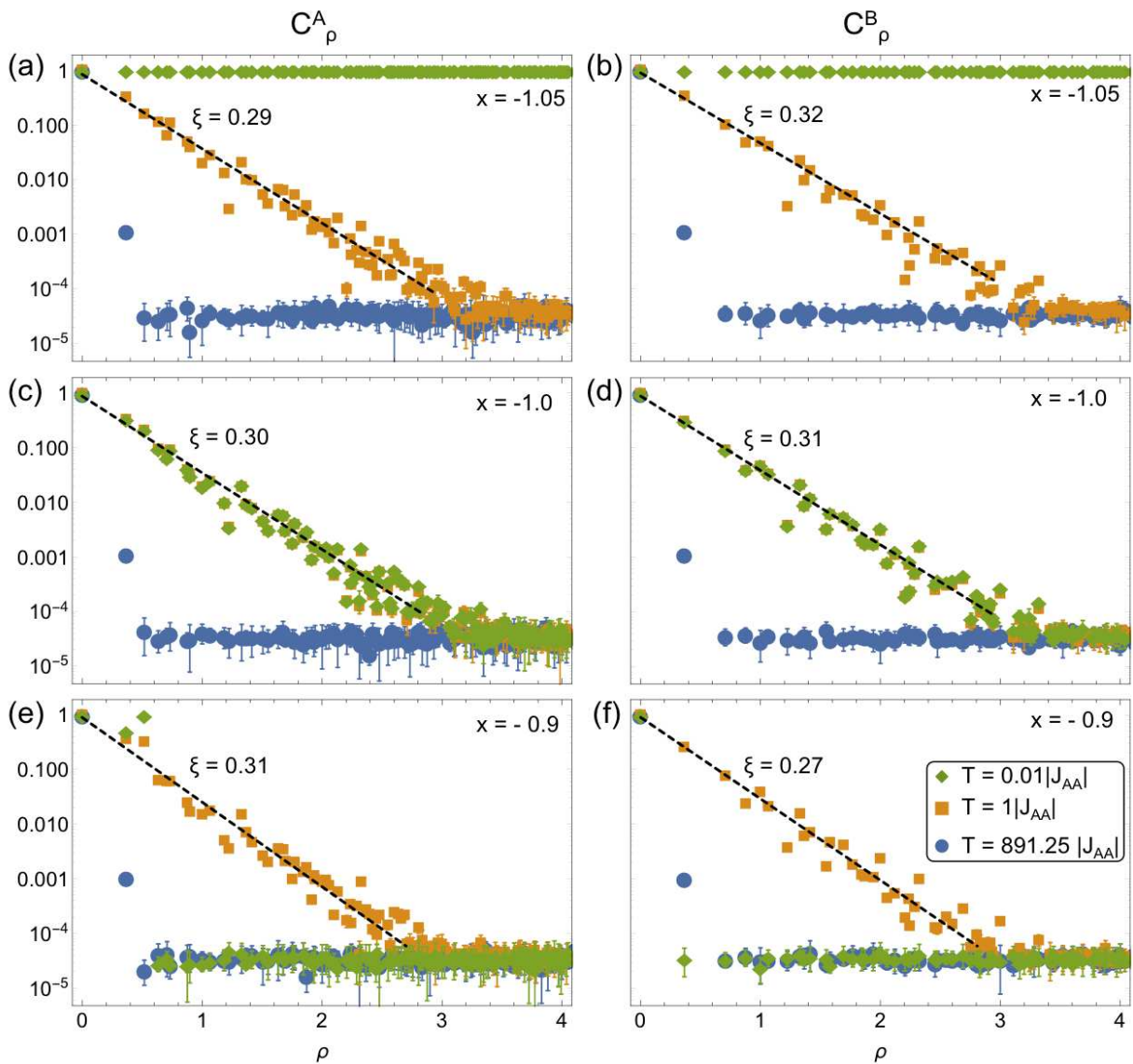}
\caption{{\bf Spin-spin correlations in the vicinity of the spin liquid phases} for $x=-0.9$ (a,b) $-1$ (c,d) and $-1.05$ (e,f), obtained from Monte Carlo simulations. The temperatures considered are $T=0.01$ (\textcolor{green}{$\blacklozenge$}), 1 (\textcolor{orange}{$\blacksquare$}) and 891.25 (\textcolor{blue}{\Large \textbullet}). Because of the anisotropy of the lattice, we want to separate correlation functions which start on A-sites (a,c,e) and B-sites (b,d,f). The radial distance is given in units of the unit-cell length. The agglomeration of data points around $C\sim 2. 10^{-5}$ is due to finite size effects. The y-axis is on a logarithmic scale.
}
\label{fig:corr}	
\end{figure*}
%%%%%%%%%%%%%%%%%%%%%

The behavior of the spin correlations in the shuriken model can be captured by calculating the correlation length between B-spins in the checkerboard model. Since $\beta \mathcal{J}_i$ is small for all of the interactions $\mathcal{J}_i$, at all temperatures $T$ (see Fig.~\ref{fig:couplings}), this can be estimated using a perturbative expansion in  $\beta \mathcal{J}_i$. For two B-spins chosen such that the shortest path between them is along nearest neighbour $\mathcal{J}_1$ bonds we obtain to leading order
\begin{eqnarray}
&&\langle \sigma^B_i \sigma^B_j \rangle= \exp\left( -\frac{r_{ij}}{\xi_{BB}} \right)
\label{eq:corrfun}
 \\
&&\xi_{BB} \approx \frac{1}{\sqrt{2}\ln\left( \frac{T}{ \mathcal{J}_1 (T)} \right)}
\label{eq:corrlen}
\end{eqnarray}
where we choose units of length such that the linear size of a unit cell is equal to 1. Details of the calculation are given in Appendix \ref{appendix:mapping}.

The correlation length between B-spins, calculated from Eq.~(\ref{eq:corrlen}), is shown for the cases $x=-0.9$ and $x=-1$ in Fig.~\ref{fig:correlationlengths}. For $x=-0.9$ the correlation length shows a non-monotonic behavior, vanishing at both high and low temperature with a maximum at $T \sim1$. On the other hand for $x=-1$, the correlation length enters a plateau for temperatures below $T\sim 1$ and the system remains in a short range correlated regime down to $T=0$. The extent of this plateau agrees with the low-temperature plateau of the reduced susceptibility in Fig.~\ref{fig:Chi_All}
%

%%%%%%%%%%%%%%%%%%%%%%%%%%%%%%%%%%%%%%%%%
%%%%%%%%%%%%%%%%%%%%%
\begin{figure*}[t]
\centering\includegraphics[width=1\textwidth]{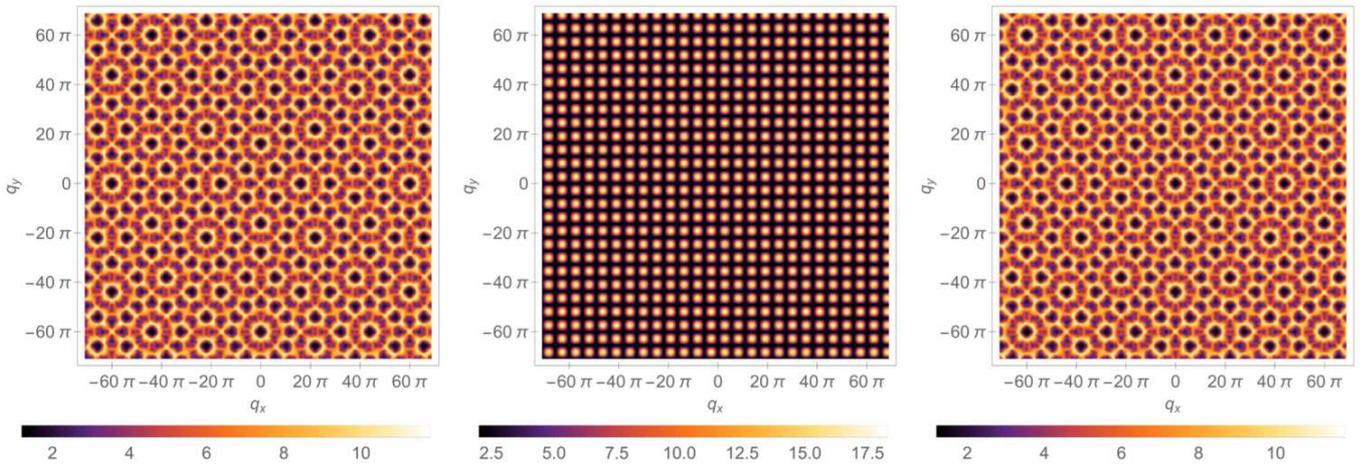}
\caption{
{\bf Static structure factors of the anisotropic shuriken lattice for} (a) $x = -1$, (b) $x = 0$ and (c) $x = 1$ at zero temperature, obtained from Monte Carlo simulations. For $x=\pm 1$, the scatterings are strongly inhomogeneous (as opposed to a standard paramagnet) and non-divergent (\textit{i.e.} without long-range order), confirming the spin liquid nature of these phases. The structure factors of the $x=+1$ and $x=-1$ models are similar by a $(q_{x},q_{y})=(60\pi,0)$ or $(0,60\pi)$ translation. The patterns are related to the 6-site unit cell of the shuriken lattice, as visible from Fig.~\ref{fig:UnitCell}. On the other hand for (b) $x=0$, the black background underlines the absence of correlations in the binary paramagnet beyond the size of the superspins (square plaquettes), which is responsible for the finite extension of the dots of scattering. In order to restore ergodicity, a local update flipping the four spins of square plaquettes was used in the simulations. A video showing the temperature dependence of the static structure factor for $x=0.9$ is available in the Supplementary Materials.
}
\label{fig:SQ}	
\end{figure*}
%%%%%%%%%%%%%%%%%%%%%
%%%%%%%%%%%%%%%%%%%%%%%%%%%%%%%%%%%%%%%%%
\subsection{Correlations and Structure factors}
\label{sec:corr}

The non-monotonic behavior of the correlation length estimated in the previous section~\ref{section:checkerboard} can be measured by Monte Carlo simulations. Let us consider the microscopic correlations both in real ($C_{\rho}$) and Fourier ($S_{q}$) space. The function $C_{\rho}$ measures the correlation between a central spin $\sigma_{0}$ and all spins at distance $\rho$. Because of the nature of the binary paramagnet, one needs to make a distinction between central spins on the A and B sublattices. Let $D_{\rho}^{X}$ be the ensemble of sites at distance $\rho$ from a given spin $\sigma_{0}^{X}$ on the $X=\{A,B\}$ sublattice. The correlation function is defined as
\begin{eqnarray}
C^X_{\rho} = \dfrac{\sum_{i\in D_{\rho}^{X}} | \langle \sigma_{0}^{X} \sigma_i \rangle |}{\sum_{i\in D_{\rho}^{X}}}
\label{eq:Crho}
\end{eqnarray}
where the absolute value accounts for the antiferromagnetic correlations. As for the static structure factor $S_{q}$, it is defined as
\begin{equation}
S_q= \langle \sigma_{\vec{q}} \,\sigma_{-\vec{q}} \rangle = \Big \langle \Big| \frac{1}{N_{uc}} \sum_i e^{-i \vec{q}\cdot \vec{r}_i} \sigma_i \Big|^2 \Big \rangle.
\end{equation}\\

$C^{A}_{\rho}$ and $C^{B}_{\rho}$ are respectively plotted on the left and right of Fig.~\ref{fig:corr}. Let us first consider what happens in absence of reentrant behavior. For $x=-1.05$ [see panels (a,b)], the system is ferromagnetic at low temperature with $C(\rho)\approx 1$ over long length scales. Above the phase transition, the correlations are exponentially decaying.

When $x=1$ [see panels (c,d)], the correlations remain exponentially decaying down to zero temperature. The correlation length $\xi$ reaches a maximum in the spin-liquid regime with $\xi \approx 0.3$. The quantitative superimposition of data for $T=0.01$ and $T=1$ is in agreement with the low-temperature plateau of the correlation length in Fig.~\ref{fig:correlationlengths}. The spin liquid remains essentially unchanged all the way up to $T\sim 1$, when defects are thermally excited. However even if the correlations are exponential, they should not be confused with paramagnetic ones, as illustrated by their strongly inhomogeneous structure factors [see Fig.~\ref{fig:SQ} and Supplementary Materials].

Once one enters the double-crossover region [see Fig.~\ref{fig:corr}.(e,f) for $x=-0.9$], the correlation function becomes non-monotonic with temperature, as predicted from the analytics of Fig.~\ref{fig:correlationlengths}. In the binary paramagnet, the B-sites are perfectly uncorrelated, while the A-sites have a finite cutoff of the correlation that is the size of the square plaquettes (superspins). This is why $S_{q}$ takes the form of an array of dots of scattering, whose width is inversely proportional to the size of the superspins [see Fig.~\ref{fig:SQ}]. Please note that the dip of correlations for the nearest-neighbors in Fig.~\ref{fig:corr}.(e) is because half of the nearest-neighbors of any A-site are on the uncorrelated B sublattice.

The intervening presence of the spin liquids between the two crossovers is conceptually reminiscent of reentrant behavior~\cite{Hablutzel39a,Vaks66a,Cladis88a}. Not in the usual sense though, since reentrance is usually considered to be a feature of ordered phases surrounded by disordered ones. But the present scenario is a direct extension of the concept of reentrance applied to disordered regimes. This reentrance is quantitatively characterized at the macroscopic level by the double-peak in the specific heat, the entropy plateau and the multi-step Curie-law crossover of Fig.~\ref{fig:MCHT}.(b), and microscopically by the non-monotonic evolution of the correlations [see Figs.~\ref{fig:correlationlengths}, \ref{fig:corr} and \ref{fig:SQ}]. As such, it provides an interesting mechanism to stabilize a gas-like phase ``below'' a spin liquid, where (a fraction of) the spins form fully correlated clusters which i) can then fluctuate independently of the other degrees-of-freedom while ii) lowering the entropy of the gas-like phase below the one of the spin liquids.

%%%%%%%%%%%%%%%%%%%%%%%%%%%%%%%%%%%%%%%%%
%%%%%%%%%%%%%%%%%%%%%%%%%%%%%%%%%%%%%%%%%
%%%%%%%%%%%%%%%%%%%%%%%%%%%%%%%%%%%%%%%%%
\section{The shuriken lattice in experiments ?}
\label{sec:exp}

Finally, we would like to briefly address the experimental situation. Unfortunately we are not aware of an experimental realization of the present model, but several directions are possible, each of them with their advantages and drawbacks.\\

The shuriken topology has been observed, albeit quite hidden, in the dysprosium aluminium garnet (DAG)~\cite{Landau71a,Wolf72a} [see Ref.~[\onlinecite{Wolf00a}] for a recent review]. The DAG material has attracted its share of attention in the 1970's, but its microscopic Hamiltonian does not respect the geometry of the shuriken lattice -- it is actually not frustrated -- and is thus quite different from the model presented in equation~(\ref{eq:ham}). However it shows that the shuriken topology can exist in solid state physics.\\

Cold atoms might offer an alternative. Indeed, the necessary experimental setup for an optical shuriken lattice has been proposed in Ref.~[\onlinecite{Glaetzle14a}]. The idea was developed in the context of spin-ice physics, \textit{i.e.} assuming an emergent Coulomb gauge theory whose intrinsic Ising degrees of freedom are somewhat different from the present model. Nonetheless, optical lattices are promising, especially if one considers that the inclusion of ``proper'' Ising spins might be available thanks to artificial gauge fields~\cite{Struck13a}.\\

But the most promising possibility might be artificial frustrated lattices, where ferromagnetic nano-islands effectively behave like Ising degrees-of-freedom. Since the early days of artificial spin ice~\cite{Wang06a}, many technological and fundamental advances have been made~\cite{Nisoli13a}. In particular, while the thermalization of the Ising-like nano-islands had been a long-standing issue, this problem is now on the way to be solved~\cite{kapaklis12a,farhan13a,farhan13b,Morgan13a,Marrows13a,Anghinolfi15a,Arnalds16a}. Furthermore, since the geometry of the nano-array can be engineered lithographically, a rich diversity of lattices is available, and the shuriken geometry should not be an issue. Concerning the Ising nature of the degrees-of-freedom, nano-islands have recently been grown with a magnetization axis $\vec z$ perpendicular to the lattice~\cite{Zhang12a,Chioar14a}.

To compute their interaction~\cite{Zhang12a,Chioar14a}, let us define the Ising magnetic moment of two different nano-islands: $\vec S = \sigma \vec z$ and $\vec S' = \sigma' \vec z$. The interaction between them is dipolar of the form
\begin{eqnarray}
D\left(\frac{\vec S\cdot \vec S'}{r^{3}}\,-\,3\frac{(\vec S\cdot \vec r)(\vec S'\cdot \vec r)}{r^{5}}\right)= \frac{D}{r^{3}}\; \sigma\,\sigma'
\label{eq:dipo}
\end{eqnarray}
where $D$ is the strength of the dipolar interaction and $\vec r$ is the vector separating the two moments. The resulting coupling is thus antiferromagnetic and quickly decays with distance. Hence, at the nearest-neighbour level, a physical distortion of the shuriken geometry -- by elongating or shortening the distance between A and B sites -- would precisely reproduce the anisotropy of equation~(\ref{eq:ham}) for $x>0$. However, the influence of interactions beyond nearest-neighbours has successively been found to be experimentally negligible~\cite{Zhang12a} and relevant~\cite{Chioar14a} on the kagome geometry. Thus the phase diagram of Fig.~\ref{fig:PhaseDiagram}.(a) could possibly be observed at finite temperature, but would likely be influenced by longer-range interactions at relatively low temperature.

%%%%%%%%%%%%%%%%%%%%%%%%%%%%%%%%%%%%%%%%%
%%%%%%%%%%%%%%%%%%%%%%%%%%%%%%%%%%%%%%%%%
%%%%%%%%%%%%%%%%%%%%%%%%%%%%%%%%%%%%%%%%%
\section{Conclusion}
\label{sec:conclusion}

The anisotropic shuriken lattice with classical Ising spins supports a variety of different phases as a function of the anisotropy parameter $x=J_{AB}/J_{AA}$: two long-range ordered ones for $|x|>1$ (ferromagnet and staggered ferromagnet) and three disordered ones [see Fig.~\ref{fig:PhaseDiagram}]. Among the latter ones, we make the distinction, at zero temperature, between two cooperative paramagnets SL$_{1,2}$ for $x=\pm 1$, and a phase that we name a binary paramagnet (BPM) for $|x|<1$. The BPM is composed of locally ordered square plaquettes separated by completely uncorrelated single spins on the B-sublattice [see Fig.~\ref{fig:PhaseDiagram}.(d)].

At finite temperature, the classical spin liquids SL$_{1,2}$ spread beyond the singular points $x=\pm 1$, giving rise to a double crossover from paramagnet to spin liquid to binary paramagnet, which can be considered as a reentrant behavior between disordered regimes. This competition is quantitatively defined by a double-peak feature in the specific heat, an entropy plateau, a multi-step Curie-law crossover and a non-monotonic evolution of the spin-spin correlation, illustrated by an inhomogeneous structure factor [see Figs.~\ref{fig:MCHT}, \ref{fig:Chi_All},\ref{fig:correlationlengths}, \ref{fig:corr} and \ref{fig:SQ}]. The reentrance can also be precisely defined by the resurgence of the couplings in the effective checkerboard model [see Fig.~\ref{fig:couplings}].\\

Beyond the physics of the shuriken lattice, the present work, and especially Fig.~\ref{fig:MCHT}, confirms the Husimi-tree approach as a versatile analytical method to investigate disordered phases such as spin liquids. Regarding classical spin liquids, Fig.~\ref{fig:Chi_All} illustrates the usefulness of the reduced susceptibility $\chi\, T$~[\onlinecite{Jaubert13a}], whose temperature evolution quantitatively describes the successive crossovers between disordered regimes. Last but not least, we hope to bring to light an interesting facet of distorted frustrated magnets, where extended regions of magnetic disorders can be stabilized by anisotropy, such as on the Cairo~\cite{Rousochatzakis12a,Rojas12a}, kagome~\cite{Li10a,Apel11a} and pyrochlore~\cite{Benton15c} lattices. Such connection is particularly promising since it expands the possibilities of experimental realizations, for example in Volborthite kagome~\cite{Hiroi01a} or breathing pyrochlores~\cite{Okamoto13a,Kimura14a}.\\

Possible extensions of the present work can take different directions. Motivated by the counter-intuitive emergence of valence-bond-crystals made of resonating loops of size 6~[\onlinecite{Ralko2015}], the combined influence of quantum dynamics, lattice anisotropy $x$~\cite{Rousochatzakis2013,Ralko2015} and entropy selection presented here should give rise to a plethora of new phases and reentrant phenomena. As an intermediary step, classical Heisenberg spins also present an extensive degeneracy at $x=1$~\cite{Richter2009,Rousochatzakis2013}, where thermal order-by-disorder is expected to play an important role in a similar way as for the parent kagome lattice, especially when tuned by anisotropy $x$. The addition of an external magnetic field~\cite{Derzhko2006,Nakano2015} would provide a direct tool to break the invariance by transformation of equation~(\ref{eq:symmetricPD}), making the phase diagram of Fig.~\ref{fig:PhaseDiagram}.(a) asymmetric. Furthermore, the diversity of spin textures presented here offers a promising framework to be probed by itinerant electrons coupled to localized spins via double-exchange.

%%%%%%%%%%%%%%%%%%%%%%%%%%%%%%%%%%%%%%%%%
\begin{acknowledgments}
We are thankful to John Chalker, Arnaud Ralko, Nic Shannon and Mathieu Taillefumier for fruitful discussions and suggestions. This work was supported by the Theory of Quantum Matter Unit of the Okinawa Institute of Science and Technology Graduate University.
\end{acknowledgments}
%%%%%%%%%%%%%%%%%%%%%%%%%%%%%%%%%%%%%%%%%
%%%%%%%%%%%%%%%%%%%%%%%%%%%%%%%%%%%%%%%%%
%%%%%%%%%%%%%%%%%%%%%%%%%%%%%%%%%%%%%%%%%
\appendix
\label{sec:appendix}

%%%%%%%%%%%%%%%%%%%%%%%%%%%%%%%%%%%%%%%%%
%%%%%%%%%%%%%%%%%%%%%%%%%%%%%%%%%%%%%%%%%
\section{Methods}
\label{sec:Methods}

Classical Monte Carlo simulations have been performed based on the single-spin-flip algorithm. Let a Monte Carlo step (MCs) be the standard Monte Carlo unit of time made of $N$ attempts to flip a spin chosen at random. Typical simulations in this paper consist of
\begin{itemize}
\item $10^{7}$ MCs, including $10^{6}$ MCs for equilibration;
\item 1 measurement every 50 MCs for $|x| > 1$ (total of $180\;000$ samples);
\item 1 measurement every 10 MCs for $|x| \leq 1$ (total of $900\;000$ samples); 
\item system sizes varying from $N =2\,400$ to $15\,000$. Fig.~\ref{fig:PhaseDiagram} has been obtained for $N =2\,400$ sites.
\end{itemize}
In order to improve the statistics, a large number of temperatures were simulated, and data were averaged over 4 neighboring temperatures.\\

To avoid any potential problems of ergodicity breaking in Monte Carlo simulations, we combined this numerical approach to analytical calculations on a Husimi tree~\cite{Husimi50a}, a method that has already demonstrated success in frustrated magnets~\cite{Chandra94a,Yoshida02a,Jaubert13a,Strecka15a}. In a nutshell, the Husimi tree is a recursive approach on a Bethe lattice where all vertices are replaced by a cluster of spins. The clusters are connected to each other via their external corners, without making any closed loops. This allows to correctly take into account the interactions within each cluster, where frustration can be encoded.

Because the shuriken lattice is made of corner-sharing triangles, a natural choice would have been to consider triangles as building blocks of the Husimi-tree recursion. However a single triangle does not properly include the geometry of the anisotropy presented in Fig.~\ref{fig:UnitCell}. This is why, in the same way as for the 16-vertex model~\cite{Foini13a,Levis13a}, we chose a larger building block made of four triangles forming a ``shuriken'' [see Fig.~\ref{fig:UnitCell}], which includes the anisotropy between A- and B-sites.

On the other hand, it neglects correlations on the length scale of the octagonal plaquettes and beyond. As such, the Husimi tree remains a mean field approximation which can only be qualitative in the vicinity of a critical point below its upper critical dimension. Since the 2D Ising Universality class is obviously not mean field, the Husimi tree underestimates the transition temperatures for $|x|>1$ by a factor of $\approx 0.7$. This is why the boundaries of the FM and SFM phases have been determined with Monte Carlo simulations [open circles in Fig.~\ref{fig:PhaseDiagram}.(a)].

But as far as disordered phases are concerned, the Husimi tree is quantitatively correct, as shown by Fig.~\ref{fig:MCHT} and table~\ref{tab:zeroT}. Being analytical, it provides an accurate way to determine the local maxima of the specific heat during crossovers [open triangles in Fig.~\ref{fig:PhaseDiagram}.(a)].

%%%%%%%%%%%%%%%%%%%%%%%%%%%%%%%%%%%%%%%%%
%%%%%%%%%%%%%%%%%%%%%
\begin{figure*}[t]
\centering\includegraphics[width=0.45 \textwidth]{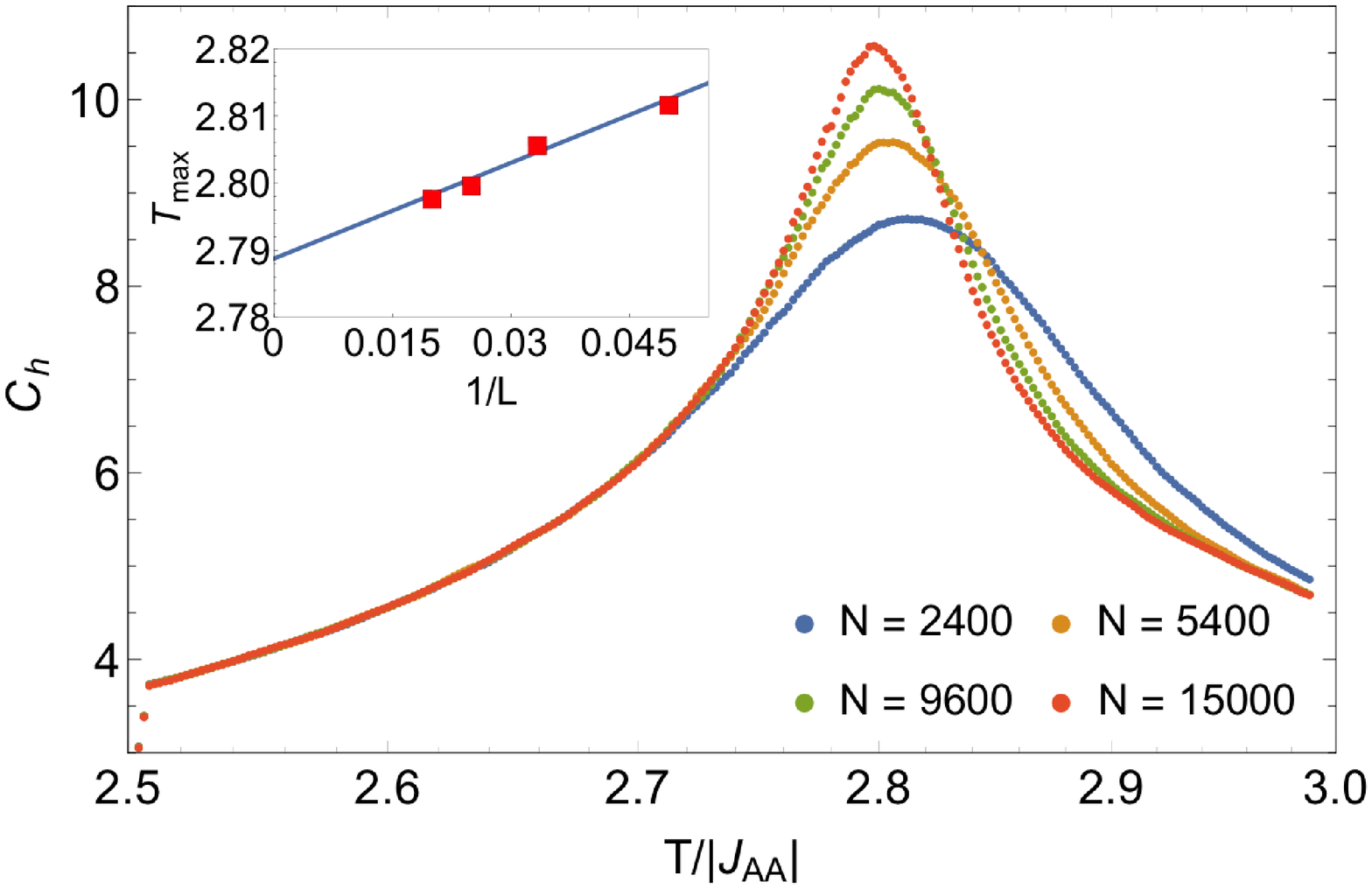}$\qquad$
\centering\includegraphics[width=0.45 \textwidth]{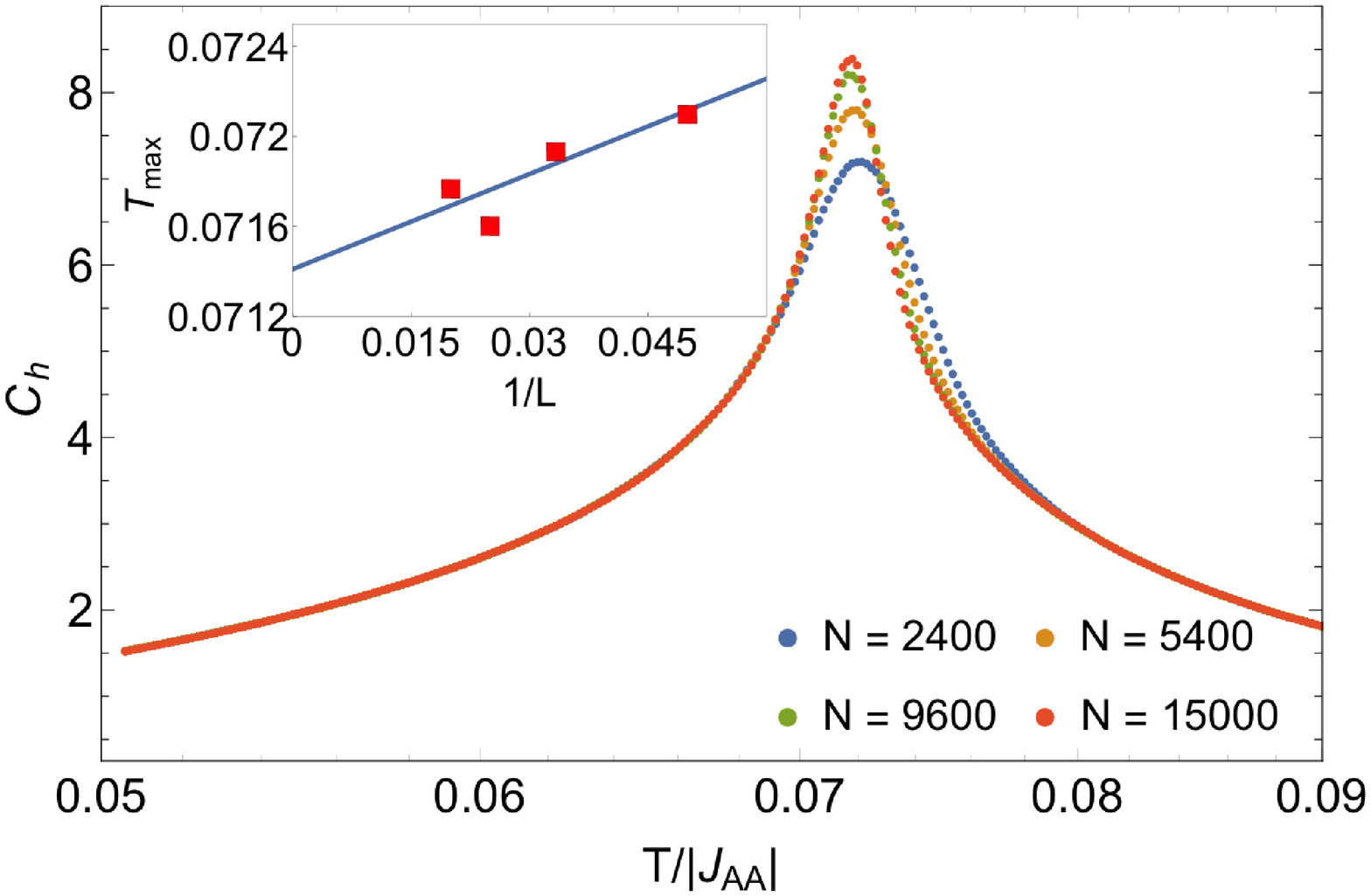}
\caption{{\bf Finite size effects on the specific heat} for $x = -3$ (\textit{left}) and $x=-1.05$ (\textit{right}). \textit{Insets:} The transition temperature is scaled as a function of $1/L$ where $L$ is the linear system size. In the thermodynamic limit, we find $T_c = 2.788(5)$ for $x=-3$ and $T_c = 0.0714(5)$ for $x=-1.05$}
\label{fig:Tc}	
\end{figure*}
%%%%%%%%%%%%%%%%%%%%%

%%%%%%%%%%%%%%%%%%%%%
\begin{figure*}
\centering\includegraphics[width=0.75 \textwidth]{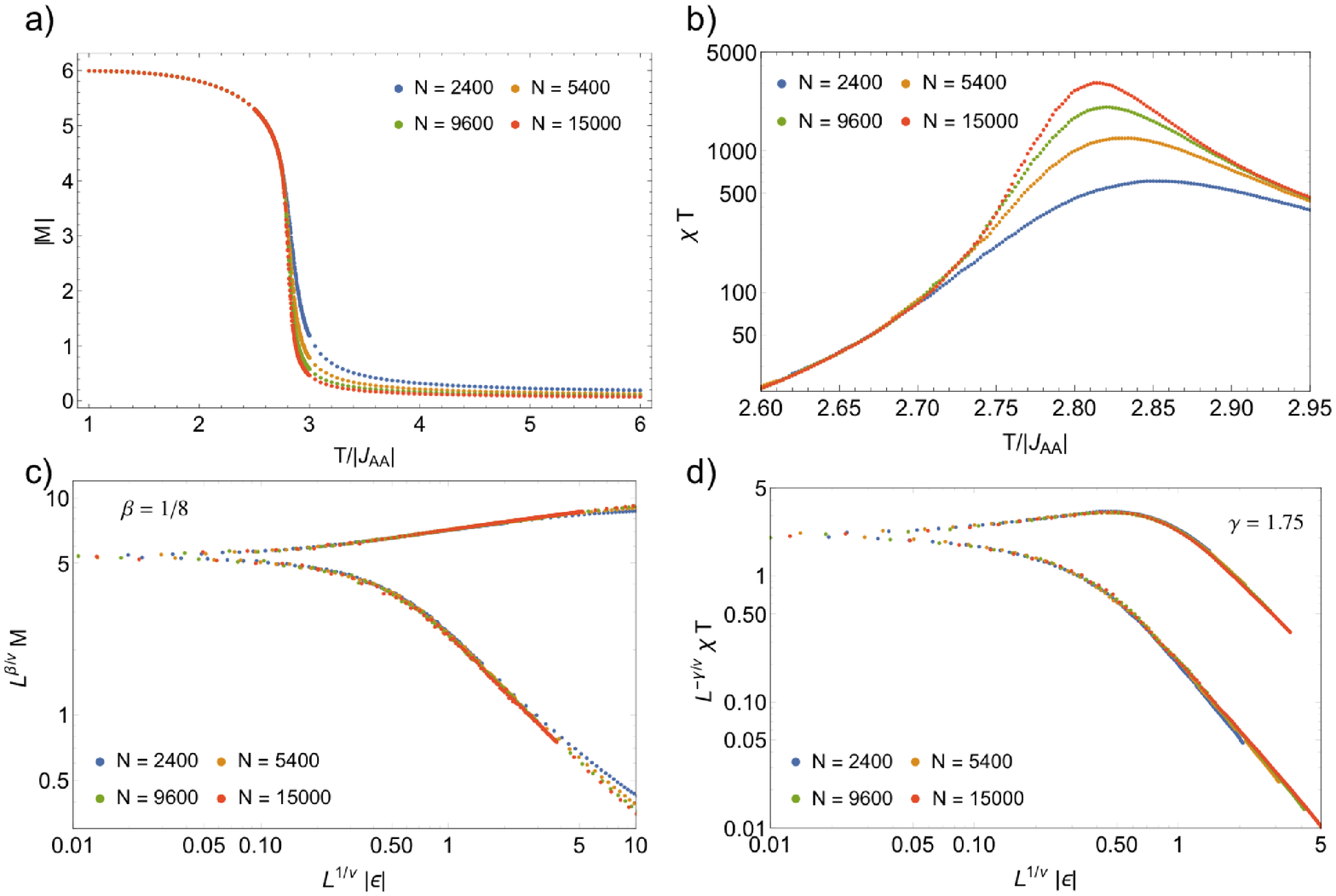}
\caption{{\bf Finite size scaling of the magnetization $|M|$ (a,c) and susceptibility $\chi$ (b,d)} for $x = -3$.}
\label{fig:CritExp1}	
\end{figure*}
%%%%%%%%%%%%%%%%%%%%%
\begin{figure*}
\centering\includegraphics[width=0.75 \textwidth]{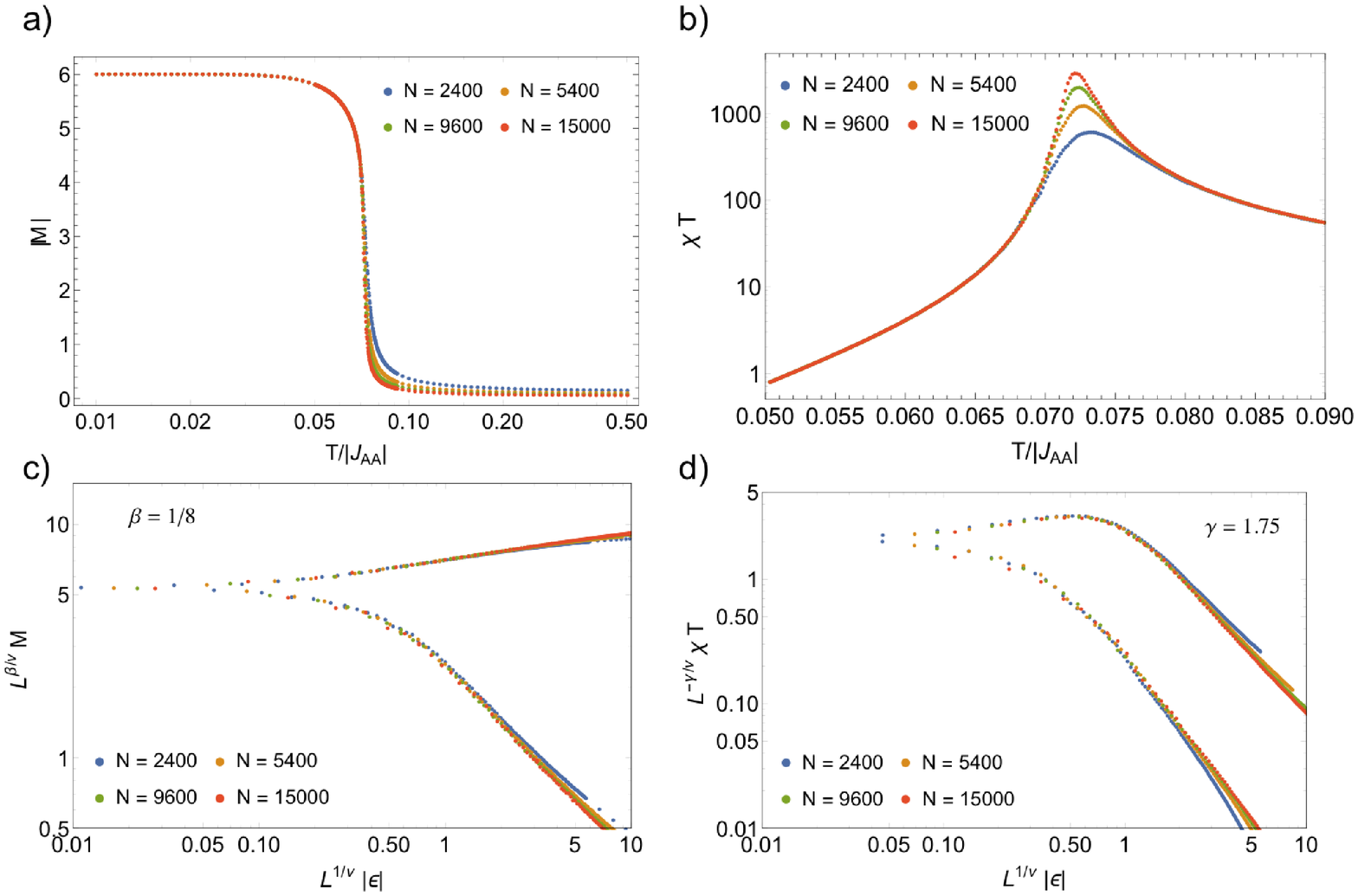}
\caption{{\bf Finite size scaling of the magnetization $|M|$ (a,c) and susceptibility $\chi$ (b,d)} for $x = -1.05$.}
\label{fig:CritExp2}	
\end{figure*}
%%%%%%%%%%%%%%%%%%%%%
%%%%%%%%%%%%%%%%%%%%%%%%%%%%%%%%%%%%%%%%%
\section{Pauling entropy of the spin liquids}

In the isotropic case ($x=1$), and by symmetry for $x=-1$ as well [see equation~(\ref{eq:symmetricPD})], a simple Pauling argument is possible for the calculation of the entropy~\cite{Pauling35a}. If $N$ is the number of Ising spins, then there are $2N/3$ triangles in the system. Out of the $2^{N}$ possible configurations, the Pauling argument states that approximately $(6/8)^{2N/3}$ are allowed in the ground state, giving a total number of ground states in the spin liquids SL$_{1,2}$
\begin{eqnarray}
\Omega_{\rm SL - Pauling}=2^{N}\left(\dfrac{6}{8}\right)^{2N/3}=\left(\dfrac{9}{2}\right)^{N/3}
\label{eq:Opauling}
\end{eqnarray}
giving an entropy
\begin{eqnarray}
S_{\rm SL - Pauling}
&=&\dfrac{N}{3}k_{B}\ln\dfrac{9}{2}\nonumber\\
&=&\dfrac{N}{6}k_{B}\ln\dfrac{40.5}{2}\nonumber\\
&\approx& N\, k_{B}\,0.50136
\label{eq:Spauling}
\end{eqnarray}

The small difference between the Pauling estimate of equation~(\ref{eq:Spauling}) and Monte Carlo results (0.50366) is mostly corrected by considering shurikens as building blocks in the Husimi-tree calculations (0.50340) [see Table~\ref{tab:zeroT}].

%%%%%%%%%%%%%%%%%%%%%%%%%%%%%%%%%%%%%%%%%
%%%%%%%%%%%%%%%%%%%%%%%%%%%%%%%%%%%%%%%%%
\section{2D Ising Universality class}
\label{sec:PhaseTransition}

For $|x| > 1$, the anisotropic shuriken model orders at low temperature via a spontaneous $\mathbb{Z}_{2}$ symmetry breaking [see Fig.~\ref{fig:PhaseDiagram}]. We know it is a critical point of the 2D Ising Universality class for large $|x|$ [see section~\ref{sec:LRO}]. In this appendix, our goal is to confirm numerically that it remains in the same Universality class as $|x|\rightarrow 1^{+}$, by considering two different values of the coupling ratio: $x = -3$ and $x = -1.05$. By symmetry of equation~(\ref{eq:symmetricPD}), the results directly apply to $x>1$ also.

In Fig.~\ref{fig:Tc}, we analyze the specific heat $C_h$ for four different system sizes $N =\{2400, 5400, 9600,15000\}$. The transition temperature scales like $1/N^{1/3}$ to its thermodynamic limit found at
\begin{eqnarray}
x=-3 &\Rightarrow& T_c = 2.788(5)\\
x=-1.05 &\Rightarrow& T_c = 0.0714(5)
\label{eq:Tc}
\end{eqnarray}

Based on these values of the transition temperature, we can define the reduced temperature $\epsilon = (T - T_c)/T_c$. Following standard finite size scaling~\cite{LandauBinder2009}, we confirm in Figs.~\ref{fig:CritExp1} and~\ref{fig:CritExp2} that the nature of the phase transition is consistent with the 2D Ising Universality class with critical exponents 
\begin{eqnarray}
\beta = 0.125,\quad \gamma = 1.75,\quad \nu = 1\\
\nonumber
\end{eqnarray}
%

%%%%%%%%%%%%%%%%%%%%%%%%%%%%%%%%%%%%%%%%%
\section{Details of the decoration-iteration transformation}
\label{appendix:mapping}

In this Appendix we give the details of the map to the effective model
on the checkerboard lattice derived in Section \ref{section:checkerboard}.
We give the derivation in Section \ref{subsec:deriv} and then give 
details of the calculation of the correlation length in Section \ref{subsec:corrlen}.
%
%\begin{figure*}
%\centering
%\includegraphics[width=0.8\textwidth]{squareZ.png}
%\caption{The partition function of the square plaquettes 
%of ${\sf a}$ spins for different, fixed, configurations
%of the surrounding B-spins [Eqs. (\ref{eq:z++++})-(\ref{eq:z+-+-})]}
%\label{fig:squarez}
%\end{figure*}

%%%%%%%%%%%%%%%%%%%%%%%%%%%%%%%%%%%%%%%%%
\subsection{Derivation of the effective model on the checkerboard lattice}
\label{subsec:deriv}

%%%%%%%%%%%%%%%%%%%%%
\begin{figure}
\centering
\includegraphics[width=0.8\columnwidth]{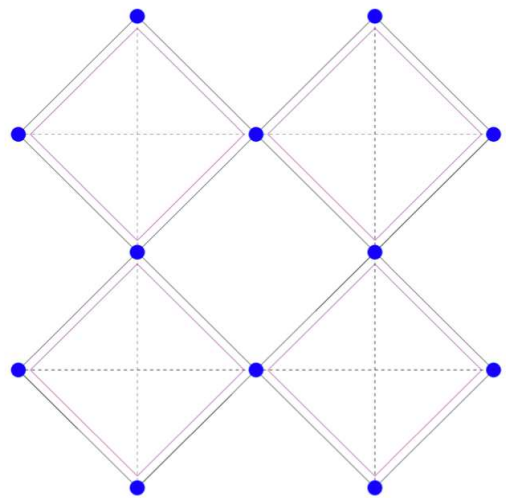}
\caption{The checkerboard lattice formed by the set of B-
spins on the shuriken lattice. %and the effective interactions
%between the B-spins
}
\label{fig:checkerboard}
\end{figure}
%%%%%%%%%%%%%%%%%%%%%

Consider the partition function of the
anisotropic shuriken model
\begin{eqnarray}
Z=
\sum_{\{ \sigma^{B}_i=\pm1\}}
\sum_{\{ \sigma^{A}_i=\pm1\}}
\exp\left[- \beta (H_{AA}+H_{AB}) \right]
\label{eq:partition1}
\end{eqnarray}
where $H_{AA}$ and $H_{AB}$ are
respectively the Hamiltonian of the square plaquettes of A-
spins
and the Hamiltonian coupling the intermediate
B-spins to the square plaquettes.
Summing over configurations of A- spins, we obtain:
\begin{eqnarray}
Z=
\sum_{\{ \sigma^{B}_i=\pm1\}}
\prod_{\square} \mathcal{Z}_{\square} (\{ \sigma^{B}_i\})
\label{eq:partition2}
\end{eqnarray}
where the product is over all the square plaquettes
of the lattice and $\mathcal{Z}_{\square} (\{ \sigma^{B}_i\})$
depends on the configuration of the four B-spins immediately neighbouring
a given square plaquette.

There are sixteen possible arrangements of the
four spins B-spins surrounding a square
plaquette of which only four are inequivalent from the point of view of
symmetry.
These  give rise to four possible values for 
$\mathcal{Z}_{\square}$:
\begin{widetext}
\begin{eqnarray}
&&\mathcal{Z}_{++++}
=2(2+4\cosh(4 \beta J_{AB})
+ \exp(-4 \beta J_{AA})
+ \exp(4 \beta J_{AA}) \cosh(8 \beta J_{AB})) 
\label{eq:z++++}
\\
&&\mathcal{Z}_{+++-}
=2(3+3\cosh(4 \beta J_{AB})+ \exp(-4 \beta J_{AA})
+ \exp(4 \beta J_{AA}) \cosh(4 \beta J_{AB})) 
\label{eq:z+++-}
\\
&&\mathcal{Z}_{++--}
=
4(1+2 \cosh(4 \beta J_{AB})+\cosh(4 \beta J_{AA}))
\label{eq:z++--}
\\
&&\mathcal{Z}_{+-+-}
=
4(3+\cosh(4 \beta J_{AA}))
\label{eq:z+-+-}
\end{eqnarray}
\end{widetext}

From these we can assign ``free energies'' $F_i=-T
\ln(\mathcal{Z}_i)$ to each of the
four possible inequivalent configurations of B-
spins around a square plaquette, i.e.
\begin{eqnarray}
F_{++++}=-T \ln(\mathcal{Z}_{++++})
\label{eq:f++++}
 \\
F_{+++-}=-T \ln(\mathcal{Z}_{+++-}) 
\label{eq:f+++-}
\\
F_{++--}=-T \ln(\mathcal{Z}_{++--})
\label{eq:f++--}
 \\
F_{+-+-}=-T \ln(\mathcal{Z}_{+-+-}) 
\label{eq:f+-+-}
\end{eqnarray}

The B-spins form a checkerboard lattice
as illustrated in Fig.~\ref{fig:checkerboard}.
Using Eqs.~(\ref{eq:z++++})-(\ref{eq:f+-+-})
we can rewrite Eq.~(\ref{eq:partition2}) in terms of an
effective Hamiltonian on the checkerboard lattice
\begin{eqnarray}
\mathcal{Z}=
\sum_{\{ \sigma^{B}_i=\pm1\}}
 \exp\left[-\beta \sum_{\boxtimes} H_{\boxtimes} \right]
\label{eq:partition3}
\end{eqnarray}
The sum $\sum_{\boxtimes}$
is a sum over the elementary units of the
checkerboard lattice.
The function $H_{\boxtimes}$
is a function only of the four B-spins around a checkerboard unit
and returns one of the four $F_i$
defined in Eqs.  (\ref{eq:f++++})-(\ref{eq:f+-+-})
as appropriate to the configuration of those four spins.

We can rewrite $H_{\boxtimes}$ explicitly in terms
of interactions between the spins on the
checkerboard lattice.
The resultant effective Hamiltonian for the spins
on the checkerboard lattice contains a
constant term $\mathcal{J}_0$,
a nearest neighbour interaction $\mathcal{J}_1$,
a second nearest neighbour interaction $\mathcal{J}_2$,
and a four-site ring interaction $\mathcal{J}_{\sf ring}$.
\begin{eqnarray}
&&H_{\boxtimes}=
-\mathcal{J}_0(T)
-\mathcal{J}_1(T) \sum_{\langle ij \rangle}\sigma^{B}_i \sigma^{B}_j  \nonumber \\
&&\quad -\mathcal{J}_2(T) \sum_{\langle \langle ij \rangle \rangle}\sigma^{B}_i \sigma^{B}_j
-\mathcal{J}_{\sf ring}(T) \prod_{i \in \boxtimes} \sigma_i^{B}
\label{eq:effectiveH}
\end{eqnarray}
All couplings are functions of temperature $\mathcal{J}_i=\mathcal{J}_i(T)$.

The relationship between the temperature dependent couplings
$\mathcal{J}_i(T)$ appearing in Eq.~(\ref{eq:effectiveH}) and 
the free energies $F_j$ defined in Eqs. (\ref{eq:f++++})-(\ref{eq:f+-+-}) is:
\begin{eqnarray}
&&\mathcal{J}_0=\frac{-1}{8}(F_{++++}+ F_{+-+-}+2F_{++--}+4F_{+++-}) 
\label{eq:J0}\nonumber \\
\\
&&\mathcal{J}_1=\frac{-1}{8}(F_{++++}-F_{+-+-}) 
\label{eq:J1}
\\
&&\mathcal{J}_2=\frac{-1}{8}(F_{++++}+F_{+-+-}- 2 F_{++--}) 
\label{eq:J2}
\\
&&\mathcal{J}_{\sf ring}=\frac{-1}{8}(F_{++++}+F_{+-+-}+2 F_{++--}-4F_{+++-})
\nonumber \\
\label{eq:Jring}
\end{eqnarray}

We have thus succeeded in mapping the original model
on the shuriken lattice, onto an effective model on the checkerboard lattice
[Eq.~(\ref{eq:effectiveH})]. 

%%%%%%%%%%%%%%%%%%%%%%%%%%%%%%%%%%%%%%%%%
\subsection{Transition temperature of the decorated square lattice}
\label{subsec:exactTc}

In the limit $x\rightarrow+\infty$, one obtains the decorated square lattice. Applying $J_{AA}=0$ to Eqs.~(\ref{eq:z++++})-(\ref{eq:f+-+-}) and then injecting the results into Eqs.~(\ref{eq:J1})-(\ref{eq:Jring}), one obtains
\begin{eqnarray}
&&\mathcal{J}_1
=\frac{1}{2\beta} \ln(\cosh(2 \beta J_{AB})),
\label{eq:bipartiteJ1}
  \\
&&\mathcal{J}_2=\mathcal{J}_{\sf ring} =0.
\end{eqnarray}
The term $\mathcal{J}_{0}$ does not cancel, but it only appears as a prefactor in the partition function of Eq.~(\ref{eq:partition3}) and thus does not influence the critical point.

Our effective model thereby becomes a square lattice with a temperature dependent nearest-neighbour coupling $\mathcal{J}_{1}(T)$. It is exactly soluble and the transition temperature $T_{c}=1/\beta_{c}$ is obtained by injecting Eq.~(\ref{eq:bipartiteJ1}) into Onsager's solution of the Ising square lattice~\cite{Onsager44a}
\begin{eqnarray}
\beta_{c} \mathcal{J}_{1}(T_{c})&=&\frac{1}{2}  \ln(\cosh(2 \beta_{c} J_{AB}))\nonumber\\
&=& \frac{1}{2} \ln(\sqrt{2}+1) \quad \textrm{(Onsager)}
\label{eq:bipartiteTc}
\end{eqnarray}
which gives the result of Eq.~(\ref{eq:exactTc})
\begin{eqnarray}
T_{c}&=&\frac{2 J_{AB}}{\ln\left(\sqrt{2}+1 +\sqrt{2+2\sqrt{2}}\right)}\nonumber\\
&\approx& 1.30841\; J_{AB}
\end{eqnarray}
%

%%%%%%%%%%%%%%%%%%%%%%%%%%%%%%%%%%%%%%%%%
\subsection{Correlation length}
\label{subsec:corrlen}

We observed in Section \ref{section:checkerboard} that for $x \leq 1$
the couplings of the effective model are small compared to the temperature,
for {\it all values of temperature}.

An expansion of the partition function of the effective model in powers
of $\beta \mathcal{J}_i$ is thus justified.
Where $|x| < 1$ this expansion is assymptotically exact in both
high and low temperature regimes.

\begin{figure}[b]
\centering\includegraphics[width=0.8\columnwidth]{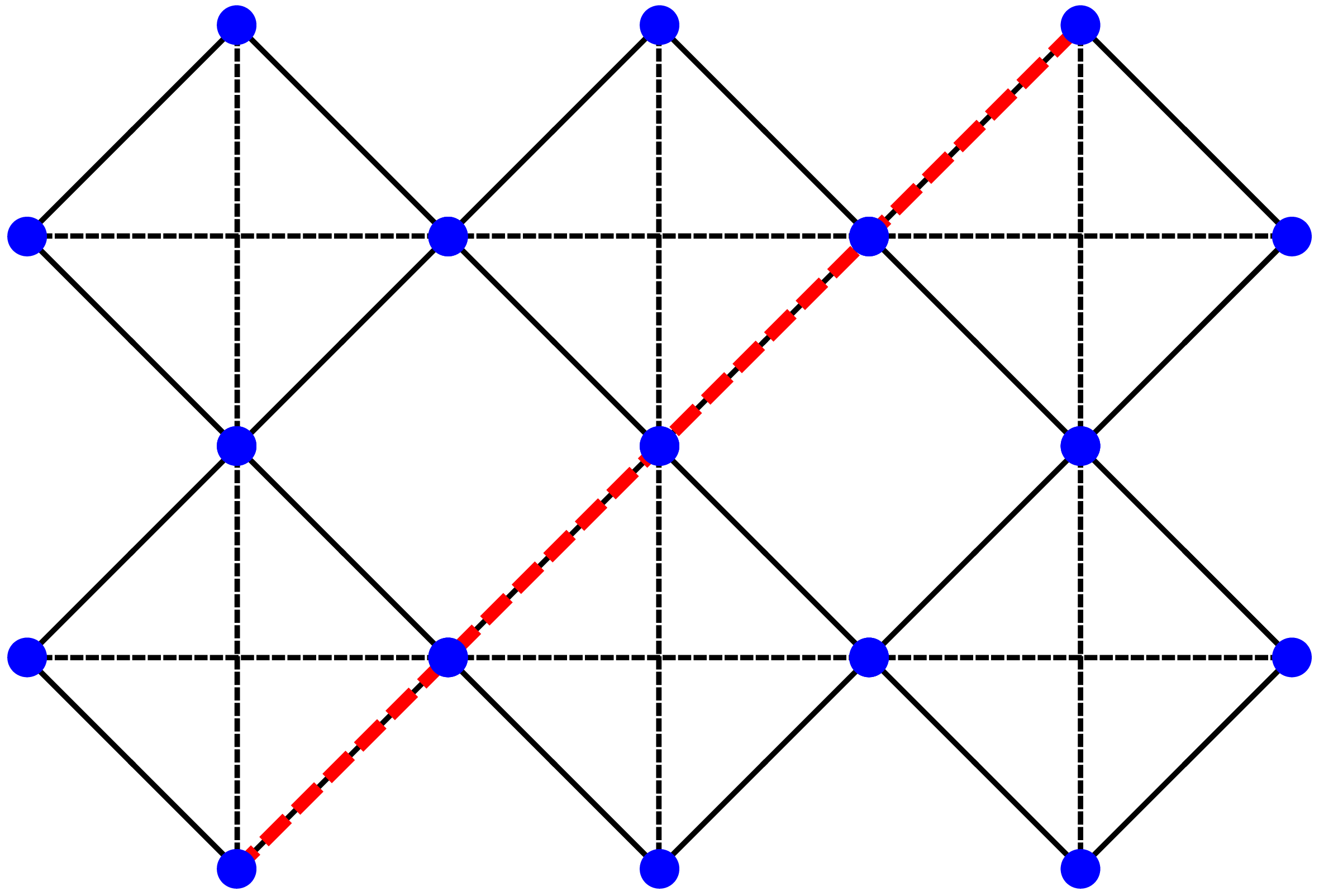}
\caption{A path (in red) between two spins on the
checkerboard lattice containing only nearest neighbour $\mathcal{J}_1$
bonds. The correlation function between two such spins in the
disordered regime is calculated in section \ref{subsec:corrlen}.}
\label{fig:separation}
\end{figure}

Here we show how to use this expansion to calculate the correlation function
$\langle \sigma^{B}_0 \sigma^{B}_{m} \rangle$ for a pair of 
B-spins.
For simplicity and concreteness we will do the calculation for a pair separated 
by a path such as that in Fig.~\ref{fig:separation}, where the shortest route between
them traverses only $\mathcal{J}_1$ bonds and contains $m$ such bonds.
However, there is no difficulty in making the calculation for other cases.

%%%%%%%%%%%%%%%%%%%%%
%%%%%%%%%%%%%%%%%%%%%

We have
\begin{widetext}
\begin{eqnarray}
&&\langle \sigma_0^B \sigma_m^B \rangle
= \nonumber \\
&&\frac{\sum_{\{ \sigma_i \pm 1 \}} 
 \sigma_0^B \sigma_m^B \exp
\left[ \beta \sum_{\boxtimes}
\mathcal{J}_0(T)+
\mathcal{J}_1(T) \sum_{\langle ij \rangle}\sigma^{B}_i \sigma^{B}_j +
\mathcal{J}_2(T) \sum_{\langle \langle ij \rangle \rangle}\sigma^{B}_i \sigma^{B}_j+
\mathcal{J}_{\sf ring}(T) \prod_{i \in \boxtimes} \sigma^{B}_i \right]}{
\sum_{\{ \sigma_i \pm 1 \}} 
 \exp\left[\beta \sum_{\boxtimes}
\mathcal{J}_0(T)+
\mathcal{J}_1(T) \sum_{\langle ij \rangle}\sigma^{B}_i \sigma^{B}_j +
\mathcal{J}_2(T) \sum_{\langle \langle ij \rangle \rangle}\sigma^{B}_i \sigma^{B}_j+
\mathcal{J}_{\sf ring}(T) \prod_{i \in \boxtimes} \sigma_i \right]
} \nonumber \\
&&=\frac{1}{N_c}\frac{\sum_{\{ \sigma_i \pm 1 \}} 
 \sigma_0^B \sigma_m^B \sum_{n=0}^{\infty}\frac{1}{n!}
\left[ \beta \sum_{\boxtimes}
\mathcal{J}_1(T) \sum_{\langle ij \rangle}\sigma^{B}_i \sigma^{B}_j +
\mathcal{J}_2(T) \sum_{\langle \langle ij \rangle \rangle}\sigma^{B}_i \sigma^{B}_j+
\mathcal{J}_{\sf ring}(T) \prod_{i \in \boxtimes} \sigma_i \right]^n}{1 + \frac{1}{N_c}
\sum_{\{ \sigma_i \pm 1 \}} 
\sum_{n=1}^{\infty}\frac{1}{n!}
\left[ \beta \sum_{\boxtimes}
\mathcal{J}_1(T) \sum_{\langle ij \rangle}\sigma^{B}_i \sigma^{B}_j +
\mathcal{J}_2(T) \sum_{\langle \langle ij \rangle \rangle}\sigma^{B}_i \sigma^{B}_j+
\mathcal{J}_{\sf ring}(T) \prod_{i \in \boxtimes} \sigma_i^{B} \right]^n
} \nonumber \\
\label{eq:HTE}
\end{eqnarray}
\end{widetext}
where $N_c$ is the total number of spin configurations of the checkerboard model.

The leading non-zero term in Eq.~(\ref{eq:HTE}) comes from the $n=m$ part of the sum in the
numerator, and corresponds to covering the shortest path between $\sigma_i^{B}$
and $\sigma_j^{B}$ with $\mathcal{J}_1$ interactions.
There are $m!$ ways of ordering the product of terms, which cancels the $\frac{1}{n!}$
occurring in the denominator.
We thus obtain
\begin{eqnarray}
\langle \sigma_i^B \sigma_j^B \rangle
\approx\left( \beta \mathcal{J}_1 (T)\right)^{m}
%=\left( \beta | \mathcal{J}_1 (T)|\right)^{\frac{r}{\sqrt{2}}}
%=\left(\frac{1}{\beta |\mathcal{J}_1 (T)|}\right)^{-{\frac{r}{\sqrt{2}}}}
=\exp\left[- m\ln\left( \frac{1}{\beta \mathcal{J}_1 (T)} \right) \right]\nonumber \\
\end{eqnarray}

In our choice of units of length, 
made such that the linear size of a unit cell equals 1,
the distance between the spins is
\begin{eqnarray}
r=\frac{m}{\sqrt{2}}
\end{eqnarray}

We therefore have a correlation length
\begin{eqnarray}
\xi_{BB}=\frac{1}{\sqrt{2}\ln\left( \frac{1}{\beta \mathcal{J}_1 (T)} \right)}
\label{eq:corrxi}
\end{eqnarray}

%%%%%%%%%%%%%%%%%%%%%%%%%%%%%%%%%%%%%%%%%
\bibliography{library}
\end{document}